%% file: paper.tex
\documentclass[conference]{IEEEtran}
\pdfoutput=1
\IEEEoverridecommandlockouts
\usepackage{cite}
\usepackage{amsmath,amssymb,amsfonts}
\usepackage{algorithmic}
\usepackage{graphicx}
\usepackage{textcomp}
\usepackage{xcolor}
\usepackage{tikz}
\usepackage{listings}
\usepackage{hyperref}
\usepackage{adjustbox}

\DeclareMathOperator{\tr}{Tr}
\usetikzlibrary{quantikz}
\graphicspath{ {./fig/} }

\begin{document}

\title{Enabling a Programming Environment for an Experimental Ion Trap Quantum Testbed}
%\thanks{Identify applicable funding agency here. If none, delete this.}}

% From Section IV-B of the IEEEtran manual
\author{
    \ifdefined\doubleblind
        \IEEEauthorblockN{Anonymous Author(s)} 
    \else
        \IEEEauthorblockN{Austin Adams\IEEEauthorrefmark{1}\IEEEauthorrefmark{2}, Elton Pinto\IEEEauthorrefmark{1}, Jeffrey Young\IEEEauthorrefmark{1}, Creston Herold\IEEEauthorrefmark{3}, \\ Alex McCaskey\IEEEauthorrefmark{4}, Eugene Dumitrescu\IEEEauthorrefmark{4}, Thomas M. Conte\IEEEauthorrefmark{1}}  
        \IEEEauthorblockA{\IEEEauthorrefmark{1}College of Computing\\ Georgia Institute of Technology, Atlanta, Georgia}
        \IEEEauthorblockA{\IEEEauthorrefmark{3}CIPHER Quantum Systems Division\\ Georgia Tech Research Institute, Atlanta, Georgia}
        \IEEEauthorblockA{\IEEEauthorrefmark{4}Computing and Computational Sciences Directorate\\ Oak Ridge National Laboratory, Oak Ridge, Tennessee}
        \IEEEauthorblockA{\IEEEauthorrefmark{2}Email: aja@gatech.edu}
    \fi
}

%\author{\IEEEauthorblockN{Austin Adams}
%\IEEEauthorblockA{\textit{College of Computing} \\
%\textit{Georgia Institute of Technology}\\
%Atlanta, GA, USA \\
%aja@gatech.edu}
%\and
%\IEEEauthorblockN{Jeff Young}
%\IEEEauthorblockA{\textit{College of Computing} \\
%\textit{Georgia Institute of Technology}\\
%Atlanta, GA, USA \\
%jyoung9@gatech.edu}
%\and
%\IEEEauthorblockN{Creston Herold}
%\IEEEauthorblockA{\textit{CIPHER Quantum Systems Division} \\
%\textit{Georgia Tech Research Institute}\\
%Atlanta, GA, USA \\
%creston.herold@gtri.gatech.edu}
%\and
%\IEEEauthorblockN{Tom Conte}
%\IEEEauthorblockA{\textit{College of Computing} \\
%\textit{Georgia Institute of Technology}\\
%Atlanta, GA, USA \\
%conte@cc.gatech.edu}
%\and
%\and

\maketitle

\input{sections/00-abstract}
\input{sections/01-intro}
\input{sections/02-background}
\input{sections/03-impl}
\input{sections/04-results}
\input{sections/05-adaption-other-hw}
\input{sections/06-rel-work}
\input{sections/07-conclusions}

\ifdefined\doubleblind
\else
    \section*{Acknowledgment}
    The first author is partially supported by the Institute for Electronics and Nanotechnology's Georgia Tech Quantum Alliance. Additionally, we acknowledge support for this work from NSF planning grant \#2016666, ``Enabling Quantum Computer Science and Engineering'' and through the ORNL STAQCS project. Finally, this research was supported in part through research infrastructure and services provided by the Rogues Gallery testbed~\cite{young:2019:rg-exp-insights} hosted by the Center for Research into Novel Computing Hierarchies (CRNCH) at Georgia Tech and funded by NSF Award Number \#2016701. We thank the anonymous reviewers for their helpful feedback.
\fi

%\begin{equation}
%a+b=\gamma\label{eq}
%\end{equation}

% \paragraph{Positioning Figures and Tables} Place figures and tables at the top and 
%Use the abbreviation ``Fig.~\ref{fig}'', even at the beginning of a sentence.

%\begin{table}[htbp]
%\caption{Table Type Styles}
%\begin{center}
%\begin{tabular}{|c|c|c|c|}
%\hline
%\textbf{Table}&\multicolumn{3}{|c|}{\textbf{Table Column Head}} \\
%\cline{2-4} 
%\textbf{Head} & \textbf{\textit{Table column subhead}}& \textbf{\textit{Subhead}}& \textbf{\textit{Subhead}} \\
%\hline
%copy& More table copy$^{\mathrm{a}}$& &  \\
%\hline
%\multicolumn{4}{l}{$^{\mathrm{a}}$Sample of a Table footnote.}
%\end{tabular}
%\label{tab1}
%\end{center}
%\end{table}

%\begin{figure}[htbp]
%\centerline{\includegraphics{fig1.png}}
%\caption{Example of a figure caption.}
%\label{fig}
%\end{figure}

%\section*{Acknowledgment}

\IEEEtriggeratref{32}
%\newpage
\bibliographystyle{IEEEtran}
\bibliography{IEEEabrv,paper}

\end{document}

%% file: sections/00-abstract.tex
\begin{abstract}

Ion trap quantum hardware promises to provide a computational advantage over classical computing for specific problem spaces while also providing an alternative hardware implementation path to cryogenic quantum systems as typified by IBM's quantum hardware. However, programming ion trap systems currently requires both strategies to mitigate high levels of noise and also tools to ease the challenge of programming these systems with pulse- or gate-level operations.

This work focuses on improving the state-of-the-art for quantum programming of ion trap testbeds through the use of a quantum language specification, QCOR, and by demonstrating multi-level optimizations at the language, intermediate representation, and hardware backend levels. We implement a new QCOR/XACC backend to target a general ion trap testbed and then demonstrate the usage of multi-level optimizations to improve circuit fidelity and to reduce gate count. These techniques include the usage of a backend-specific numerical optimizer and physical gate optimizations to minimize the number of native instructions sent to the hardware. We evaluate our compiler backend using several QCOR benchmark programs, finding that on present testbed hardware, our compiler backend maintains the number of two-qubit native operations but decreases the number of single-qubit native operations by 1.54 times compared to the previous compiler regime. For projected testbed hardware upgrades, our compiler sees a reduction in two-qubit native operations by 2.40 times and one-qubit native operations by 6.13 times. 
\end{abstract}

\begin{IEEEkeywords}
compilation, multi-level optimization, quantum computing, ion trap
\end{IEEEkeywords}

%% file: sections/01-intro.tex
\section{Introduction}
Quantum computing promises new computational capabilities over classical computing with quantum algorithms having a theoretical exponential speedup over some classical algorithms \cite{nielsen_quantum_2010}. However, given the high error rates of present qubits (quantum bits), computational capability today reaches only as far as is achievable on NISQ (Noisy Intermediate Scale Quantum) devices, near-term quantum computers with 50-100 qubits which provide highly noisy output \cite{preskill_quantum_2018}. Limited coherence time and high gate error rates require compilers for NISQ systems to minimize the number of quantum gates (instructions) and to embrace error mitigation techniques to increase the likelihood of useful results \cite{chong_programming_2017}.

In this work, we detail our implementation of an XACC~\cite{mccaskey_xacc_2020} compiler backend targeting an experimental quantum testbed hosted by the GTRI (Georgia Tech Research Institute) Quantum Systems Division \cite{herold_universal_2016}. This new compiler backend provides a hardware-agnostic programmer-driven flow for programming the testbed using quantum circuits written inline in C++ via QCOR\cite{mintz_qcor_2020}. Our new QCOR-based implementation provides a contrast with the existing testbed tooling, which is based on proprietary software more oriented towards hardware experts. Moreover, by integrating with QCOR, programmers gain access to existing QCOR tooling, such as quantum assembly parsers, circuit optimizers, a variational workflow, and a standard library of quantum subroutines.

%To achieve these goals, the current work provides the following contributions: 
The current work provides the following contributions:
%\aja{originally we said ``To achieve these goals,'' but GTRI reviewer pointed out it wasn't clear what those goals are. So I removed that part, not sure if it needs to be replaced with something else}
\begin{itemize}
\item We demonstrate a new ion trap backend for XACC that interacts with the low-level capture system used to program and interact with a target ion trap testbed.
%\item Through the use of multi-level optimization with QCOR and XACC, we show how optimizations for a quantum program can be implemented at the language level \aja{are we referring to at the C level here? maybe just say at the hardware-agnostic (QCOR PassManager) and hardware-specific levels}, intermediate representation (IR) transformation level, and for specific hardware backends.
\item Through the use of multi-level optimization with QCOR and XACC, we show how optimizations for a quantum program can be implemented at the hardware-agnostic level by QCOR and at the hardware-specific level in particular backends.
\item We explore the usage of these optimizations to improve circuit fidelity and reduce native gate count.
\item We investigate native gate count reduction achievable with future hardware upgrades.
%\item We investigate how the migration of optimizations from the low-level hardware backend to XACC can provide scalability improvements. \aja{I don't think we talk about this, probably remove this bullet(?)}
\end{itemize}

To evaluate our work, we run a set of QCOR example programs against the backend, with the existing control code configured to respond with simulation results. We also compare gate count with the existing compiler for the testbed.
%as well as demonstrate the ease of writing the programs in QCOR versus in the original assembly format \aja{We may not address this}.

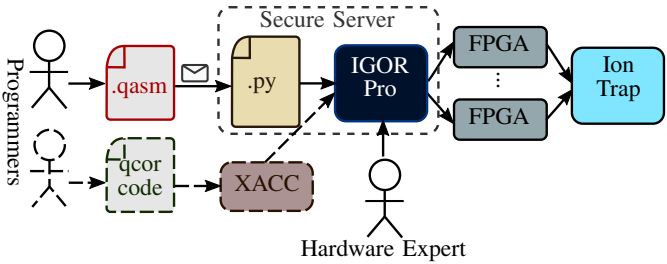
\begin{figure}
    {\def\svgwidth{\linewidth}\small\input{fig/flow-diagram-text}}\caption{Flow for programming the GTRI quantum testbed. The dashed path starting from the bottom left represents our contribution.}\label{fig:flow}
\end{figure}

\subsection{Motivation for a New Ion Trap Backend}\label{sec:motivation}
%\ch{This section provided the "why" I was lacking when reading the previous section. Are the sections dictated by IEEE? Or can the Introduction end with this paragraph, followed by background details, then implementation?}
The current approach to interacting with the GTRI testbed requires using IGOR Pro (a programming environment focused on data visualization) and detailed knowledge of the testbed mechanics --- generally, it is hardware expert--driven. An email-based submission scheme shown in Fig.~\ref{fig:flow} improved this situation by allowing for submission and parsing of quantum assembly input files, but quantum assembly--level programming is neither particularly strong as a programming environment nor suitable for near-term heterogeneous quantum--classical computing. We believe that connecting QCOR to the testbed introduces a more programmer-driven flow, making the GTRI testbed more accessible to a wider audience of quantum and classical programmers who may be unaware of hardware details. To further ensure its usefulness, we attempt to optimize programmer circuits into as few noisy primitive operations as possible. We describe our backend implementation and characterize its performance in the following sections.

%The current approach to interacting with the GTRI testbed requires IGOR Pro and detailed knowledge of the testbed mechanics --- generally, it is ``expert-driven.'' The email-based submission scheme shown in Fig.~\ref{fig:flow} improved this situation, and we even used it while active, but email is neither particularly strong as a programming environment nor suitable for near-term heterogeneous quantum--classical computing. We believe that connecting QCOR to the testbed introduces a more ``programmer-driven'' flow, making the GTRI testbed more accessible to a wider audience of NISQ programmers.

%Why is GTRI system expert-driven?
%
%Program at physical gate level - move to logical level with XACC
%* Paper contributes link and optimizations
%
%How do we address the diversity of backends?
%
%Pi/2 rotations are hard to implement at low level and are tied to GTRI testbed
%* Move up the stack and make available for other ion trap systems
%
%QCOR might let you express more than Igor "supported assembly".
%* Can take advantage of existing optimizers in XACC - what are they?

%% file: fig/flow-diagram-text.tex
%% Creator: Inkscape 1.1 (1:1.1+202106031931+af4d65493e), www.inkscape.org
%% Accompanies image file 'flow-diagram.pdf' (pdf, eps, ps)
\begingroup%
  \makeatletter%
  \providecommand\color[2][]{%
    \errmessage{(Inkscape) Color is used for the text in Inkscape, but the package 'color.sty' is not loaded}%
    \renewcommand\color[2][]{}%
  }%
  \providecommand\transparent[1]{%
    \errmessage{(Inkscape) Transparency is used (non-zero) for the text in Inkscape, but the package 'transparent.sty' is not loaded}%
    \renewcommand\transparent[1]{}%
  }%
  \providecommand\rotatebox[2]{#2}%
  \newcommand*\fsize{\dimexpr\f@size pt\relax}%
  \newcommand*\lineheight[1]{\fontsize{\fsize}{#1\fsize}\selectfont}%
  \ifx\svgwidth\undefined%
    \setlength{\unitlength}{846.89862647bp}%
    \ifx\svgscale\undefined%
      \relax%
    \else%
      \setlength{\unitlength}{\unitlength * \real{\svgscale}}%
    \fi%
  \else%
    \setlength{\unitlength}{\svgwidth}%
  \fi%
  \global\let\svgwidth\undefined%
  \global\let\svgscale\undefined%
  \makeatother%
  \begin{picture}(1,0.38830452)%
    \lineheight{1}%
    \setlength\tabcolsep{0pt}%
    \put(0,0){\includegraphics[width=\unitlength,page=1]{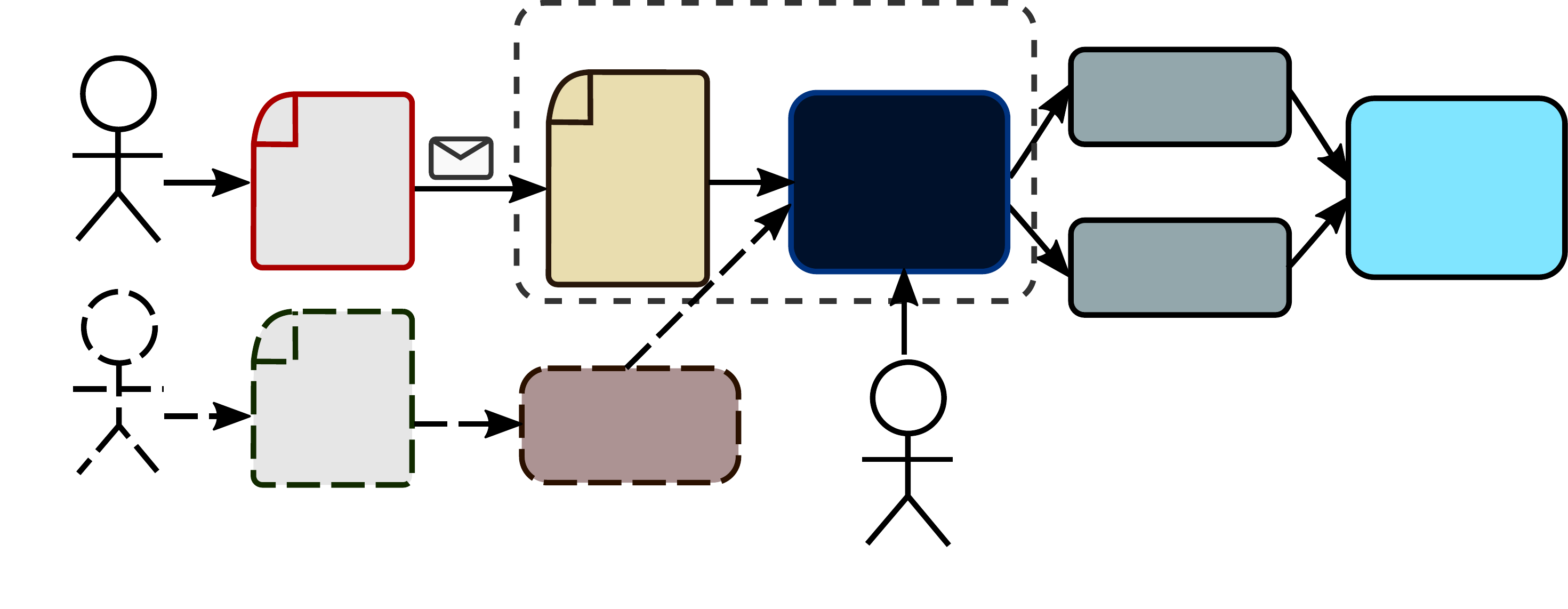}}%
    \put(0.87490339,0.31281816){\color[rgb]{0,0,0}\makebox(0,0)[lt]{\begin{minipage}{0.11048008\unitlength}\centering Ion\\ Trap\end{minipage}}}%
    \put(0.6958367,0.34493542){\color[rgb]{0,0,0}\makebox(0,0)[lt]{\begin{minipage}{0.11048008\unitlength}\centering FPGA\end{minipage}}}%
    \put(0.6958367,0.23611246){\color[rgb]{0,0,0}\makebox(0,0)[lt]{\begin{minipage}{0.11048008\unitlength}\centering FPGA\end{minipage}}}%
    \put(0.4521593,0.03723589){\color[rgb]{0,0,0}\makebox(0,0)[lt]{\begin{minipage}{0.24982743\unitlength}\centering Hardware Expert\end{minipage}}}%
    \put(0.03723589,0.34832675){\color[rgb]{0,0,0}\rotatebox{-90}{\makebox(0,0)[lt]{\begin{minipage}{0.268192\unitlength}\centering Programmers\end{minipage}}}}%
    \put(0.5171487,0.31277263){\color[rgb]{1,1,1}\makebox(0,0)[lt]{\begin{minipage}{0.11048008\unitlength}\centering IGOR\\ Pro\end{minipage}}}%
    \put(0.3724736,0.26521042){\color[rgb]{0.15686275,0.09019608,0.04313725}\makebox(0,0)[lt]{\lineheight{1.25}\smash{\begin{tabular}[t]{l}.py\end{tabular}}}}%
    \put(0.49290534,0.356688){\color[rgb]{0.2,0.2,0.2}\makebox(0,0)[t]{\lineheight{1.25}\smash{\begin{tabular}[t]{c}Secure Server\end{tabular}}}}%
    \put(0.34711885,0.13544807){\color[rgb]{0.16862745,0.06666667,0}\makebox(0,0)[lt]{\begin{minipage}{0.11048008\unitlength}\centering XACC\end{minipage}}}%
    \put(0.16419502,0.15596745){\color[rgb]{0.06666667,0.16862745,0}\makebox(0,0)[lt]{\begin{minipage}{0.09597567\unitlength}\centering qcor\\ code\end{minipage}}}%
    \put(0.16856737,0.25732978){\color[rgb]{0.66666667,0,0}\makebox(0,0)[lt]{\lineheight{1.25}\smash{\begin{tabular}[t]{l}.qasm\end{tabular}}}}%
    \put(0.74766353,0.29054189){\color[rgb]{0,0,0}\rotatebox{-90}{\makebox(0,0)[lt]{\lineheight{1.25}\smash{\begin{tabular}[t]{l}...\end{tabular}}}}}%
  \end{picture}%
\endgroup%

%% file: sections/02-background.tex
\section{Background}\label{sec:background}

\subsection{Ion Trap Quantum Computers}\label{sec:iontrap}
Quantum computers based on an ion trap realize qubits by manipulating the internal spin-like degrees of ``trapped'' atomic ions with electromagnetic radiation~\cite{nielsen_quantum_2010}. Native operations can be broadly categorized as either single- or multi-qubit operations which may comprise different universal gate sets. For example, single qubit operations include $R_\phi(\theta)$, a rotation of $\theta$ around an axis $\phi$ in the equatorial plane of the Bloch sphere, shown in~(\ref{eq:rot-exp}-\ref{eq:rot-mat})~\cite{maslov_basic_2017}. %\ch{(3) doesn't follow from (2) and $R_Z, R_X$ not defined. Will you use this later?}.
For the multi-qubit entangling operation we consider the M{\o}lmer-S{\o}renson (MS) interaction \cite{molmer_multiparticle_1999,sorensen_quantum_1999}, which defines the MS gate in (\ref{eq:ms-exp}-\ref{eq:ms-mat}).
\begin{align}
\sigma_\phi &= (\cos\phi)\sigma_x + (\sin\phi)\sigma_y \\
R_\phi(\theta) &= \exp(-i\sigma_\phi\theta/2) \label{eq:rot-exp} \\
               %&= R_Z(\phi)R_X(\theta)R_Z(-\phi) \\
               &= \begin{bmatrix}\cos \theta/2 & -ie^{-i\phi}\sin\theta/2 \\
                  -ie^{i\phi}\sin\theta/2 & \cos \theta/2
                  \end{bmatrix} \label{eq:rot-mat} \\
%MS(\alpha,\phi_L,\phi_R) &= \exp(-i\alpha(\sigma_{\phi_L} \otimes \sigma_{\phi_R})) \label{eq:xx-gate-eq} \\
\beta_{\ell r} &= -ie^{i((-1)^\ell\phi_L + (-1)^r\phi_R)}\sin\alpha \\
MS(\alpha) &= \exp(-i\alpha(\sigma_{\phi_L} \otimes \sigma_{\phi_R})) \label{eq:ms-exp} \\
           &= \begin{bmatrix}
\cos\alpha & 0 & 0 & \beta_{11} \\
0 & \cos\alpha & \beta_{10} & 0 \\
0 & \beta_{01} & \cos\alpha & 0 \\
\beta_{00} & 0 & 0 & \cos\alpha
\end{bmatrix} \label{eq:ms-mat}
\end{align}

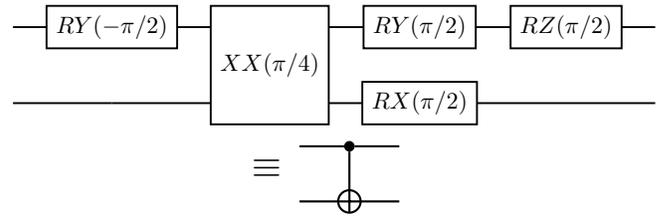
\begin{figure}
%\begin{center}
\centering
    %%%\begin{tikzpicture}
    %%%    \node[scale=0.9]{
    \begin{adjustbox}{width=\linewidth}
            \begin{quantikz}
            \qw & \gate{RY(-\pi/2)} & \gate[2]{XX(\pi/4)} & \gate{RY(\pi/2)} & \gate{RZ(\pi/2)} & \qw \\
            \qw & \qw & & \gate{RX(\pi/2)}  & \qw & \qw
            \end{quantikz}
            \end{adjustbox}
            %& \midstick[2,brackets=none]{$\equiv$}& \ctrl{1} & \qw & & \\
            %&                                     & \targ{}  & \qw & &
            %\begin{quantikz}
            %\qw & \gate{R_y(-\pi/2)} & \gate[2]{XX(\pi/4)} & \gate{R_y(\pi/2)} & \gate{R_z(\pi/2)} & \midstick[2,brackets=none]{$\equiv$}\qw& \ctrl{1} & \qw \\
            %\qw & \qw & & \gate{R_x(\pi/2)}  & \qw               & \qw & \targ{} & \qw
            %\end{quantikz}
            %\begin{quantikz}
            %\qw & \gate{R_y(-\pi/2)} & \gate{R_z(\phi_L)}\gategroup[wires=2,steps=3]{$XX(\pi/4)$} & \gate[2]{MS(\pi/4)} & \gate{R_z(-\phi_L)} & \gate{R_y(\pi/2)} & \gate{R_z(\pi/2)} & \midstick[2,brackets=none]{$\equiv$}\qw& \ctrl{1} & \qw \\
            %\qw & \qw & \gate{R_z(\phi_R)} & & \gate{R_z(-\phi_R)} & \gate{R_x(\pi/2)}  & \qw               & \qw & \targ{} & \qw
            %\end{quantikz}
            %$\equiv$\begin{quantikz}
            %%\qw\gategroup[wires=2,steps=2,style={opacity=0}]{} & \ctrl{1} & \ghost[2]{MS(\pi/4)}\qw \\
            %\qw\gategroup[wires=2,steps=3,style={opacity=0}]{} & \ctrl{1} & \gate[2]{MS(\pi/4)}\qw \\
            %\qw & \targ{} & \qw
            %\end{quantikz}
    %%%    };
    %%%\end{tikzpicture}
    {\Large $\equiv$}\begin{quantikz}
    \qw & \ctrl{1} & \qw \\
    \qw & \targ{}  & \qw 
    \end{quantikz}
%\end{center}
%\caption{CNOT decomposition. This circuit performs a CNOT up to a global phase $e^{-i\pi/4}$. \revone{The authors could improve the clarity of the introduction by being a little more explicit in the description of figure 2. The inclusion of a table with the ``typical'' CNOT operation as well as the optimized CNOT operation to go along with the schematic diagram would be helpful to explain the differences that are being implemented.}
\caption{Implementation of CNOT on the testbed using single-qubit gates and the native $XX(\pi/4)$ entangling gate. The circuit shown is equal to a CNOT up to an unimportant global phase $e^{-i\pi/4}$. %\revone{The authors could improve the clarity of the introduction by being a little more explicit in the description of figure 2. The inclusion of a table with the ``typical'' CNOT operation as well as the optimized CNOT operation to go along with the schematic diagram would be helpful to explain the differences that are being implemented.}\aja{Tried to fix}
%Here we assume the control qubit is the left ion, but $\phi_L$ and $\phi_R$ should be swapped if not.
}\label{fig:cnot-decomp}
\end{figure}

When the MS phase angles for the left and right qubits, $\phi_L$ and $\phi_R$ respectively, are zero, we have $\sigma_{\phi_L} = \sigma_{\phi_R} = \sigma_x$, yielding the $XX$-Ising gate shown in (\ref{eq:xx-gate-mat}). For simplicity, we will use this convention for the rest of this work. The $XX$ gate can be used to realize a CNOT operation when combined with single qubit gates such as those shown in Fig.~\ref{fig:cnot-decomp}~\cite{debnath_demonstration_2016,maslov_basic_2017}.
%\ch{It'd make more sense if figure was at top of this page instead of the table.}
\begin{align}
XX(\alpha) &= \begin{bmatrix}
\cos\alpha & 0 & 0 & -i\sin\alpha \\
0 & \cos\alpha & -i\sin\alpha & 0 \\
0 & -i\sin\alpha & \cos\alpha & 0 \\
-i\sin\alpha & 0 & 0 & \cos\alpha
\end{bmatrix} \label{eq:xx-gate-mat}
\end{align}

Together, arbitrary single-qubit gates provided by $R_\phi(\theta)$~\cite{maslov_basic_2017} and the $XX$ entangling gate support universal quantum computation \cite{brylinski_universal_2001,bremner_practical_2002}. Additionally, ion trap systems can support all-to-all qubit connectivity and parallel gate execution using tightly-focused individual gate beams \cite{figgatt_parallel_2019} or ion transport with stationary beams \cite{de_clercq_parallel_2016,pino_demonstration_2021}.

%\begin{equation}
%\begin{bmatrix}
%\cos\alpha & 0 & 0 & -ie^{i\phi}\sin\alpha \\
%0 & \cos\alpha & -ie^{i\phi}\sin\alpha & 0 \\
%0 & -ie^{i\phi}\sin\alpha & \cos\alpha & 0 \\
%-ie^{i\phi}\sin\alpha & 0 & 0 & \cos\alpha
%\end{bmatrix} \label{eq:xx-gate-mat}
%\end{equation}

\subsection{GTRI Quantum Testbed}\label{sec:gtri-testbed}
Researchers at GTRI have built a quantum testbed based on an ion trap~\cite{herold_universal_2016,fallek_transport_2016}. The physical apparatus consists of a stationary set of lasers which operate on ions (qubits) in the chain. The chain itself is transported to allow the lasers to target different qubits~\cite{herold_universal_2016}. Originally, the gate laser beams always targeted two ions simultaneously, but due to recent equipment upgrades allowing gate beams to target individual qubits, we assume single-qubit addressing in this work.
The testbed does not have a tightly-focused laser beam for each ion, so we assume nearest-neighbor connectivity for two-qubit gates and serialized (i.e., no parallel) single-qubit gates.
%However, we have implemented support in our compiler for  hardware upgrades.
%Despite the possibility of parallelized single-qubit gates as a result of single-qubit addressing, we produce a serialized sequence of native operations as the control software expects.
%allowing for parallelized single-qubit gates in this work.
%Additionally, we continue to assume nearest-neighbor connectivity between ions in the chain for two-qubit gates, although this does not affect our design whatsoever since QCOR itself handles qubit placement.

When configured as a general-purpose quantum computer as we assume in this work, the testbed has the $XX(\pi/4)$ gate and, for ease of calibration, the subset of $R_\phi(\theta)$ gates with $\theta = \pi/2$ as its native operations.
%Equations (\ref{eq:pirot-def}-\ref{eq:pirot-matrix}) show the definition of these $\pi/2$ rotation gates \ch{Why not directly from (4)?}.
%\begin{align}
%R_\phi(\pi/2) &= R_Z(\phi)R_X(\pi/2)R_Z(-\phi) \label{eq:pirot-def}  \\
%&= \frac{1}{\sqrt{2}}\begin{bmatrix}
%1 & -ie^{-i\phi} \\
%-ie^{i\phi} & 1 \\
%\end{bmatrix} \label{eq:pirot-matrix}
%%&= \frac{1}{\sqrt{2}}\begin{bmatrix}
%%1 & -\sin\phi - i\cos\phi \\
%%\sin\phi - i\cos\phi & 1 \\
%%\end{bmatrix} \label{eq:pirot-matrix}
%\end{align}
%Control code for the testbed is written in procedures for IGOR Pro~\cite{igorpro}, a graphical data visualization and experimentation package similar to Matlab. \ch{Is it typical to give specific company names in IEEE? Not sure what it's adding. Either way, we should at least say IGOR-based compiler or come up with another name.}
The control software for the testbed includes a rudimentary compiler that converts a dialect of quantum assembly to a sequence of native operations ($XX(\pi/4)$ and $R_\phi(\pi/2)$) according to which the control software programs FPGAs as needed to run the circuit on hardware. An example of the sequence for a Bell state circuit is shown in Table \ref{tab:seq-ex}. The existing compiler decomposes CNOTs as shown in Fig.~\ref{fig:cnot-decomp}, and it applies arbitrary single-qubit unitaries using an average of 3.25 primitive $R_\phi(\pi/2)$ rotations~\cite{herold_universal_2016}. 

\begin{table}[tbp]
\caption{Sequence of Native Operations Produced by the Existing Compiler for the Bell State Circuit \texttt{H 0; CNOT 0,1}.}\label{tab:seq-ex}
\begin{center}
\begin{tabular}{lll}
\hline
Operation & Target Ion & $\phi$ \\ \hline
$R_\phi(\pi/2)$	& 0	& $-\pi/2$ \\
$R_\phi(\pi/2)$	& 0	& $-\pi/2$ \\
$R_\phi(\pi/2)$	& 0	& $-\pi$ \\
%$R_\phi(\pi/2)$	& 0	& $\pi$ \\
$R_\phi(\pi/2)$	& 0	& $-\pi$ \\
$XX(\pi/4)$	& 0,1	& N/A \\
$R\phi(\pi/2)$	& 0	& $\pi/2$ \\
$R\phi(\pi/2)$	& 1	& 0 \\ \hline
%$R\phi(\pi/2)$	& 1	& 0 \\
%$R\phi(\pi/2)$	& 0	& $\pi$ \\
%$R\phi(\pi/2)$	& 0	& $\pi/2$ \\ \hline
%\multicolumn{4}{l}{$^{\mathrm{a}}$Sample of a Table footnote.}
\end{tabular}
\end{center}
\end{table}

%\aja{removed a strange paragraph here about the email submission script}
%Until the testbed hardware was recently repurposed, a Python script received quantum assembly by email, wrote it to a directory monitored by an IGOR script, and then read measurements written by the IGOR script to a specific directory. The script then emailed the result back to the user. This flow is shown as the upper path of Fig.~\ref{fig:flow}. \ch{this paragraph felt out of place. What's it's purpose?}

\subsection{QCOR and XACC}

Oak Ridge National Laboratory has developed the QCOR compiler specification~\cite{mintz_qcor_2020} and a reference implementation~\cite{nguyen_extending_2020}.
%XACC compiler framework as part of the QCAT (Quantum Computing Applications Teams) project.
Both aim to accelerate the development of new applications on NISQ hardware by providing a unified, automated software stack for writing quantum algorithms and mapping them to hybrid classical--quantum systems, such as a CPU-based server paired with a quantum accelerator~\cite{qcor_website}. In particular, the QCOR implementation integrates with Clang to allow writing quantum kernels inline in C++ similarly to CUDA kernels, as shown in Listing \ref{lst:qcor-ex}.

\begin{lstlisting}[basicstyle=\ttfamily,caption={Example C++ program using QCOR to generate and execute a Bernstein--Vazirani circuit for a user-provided secret bitstring.},label=lst:qcor-ex]
__qpu__ void bernstein_vazirani(qreg q,
        std::string &secret_bits) {
    int n = secret_bits.size();
    // prepare ancilla in |1>
    X(q[n]);
    // input superpositions
    H(q);
    // oracle
    for (int i = 0; i < n; i++) {
        if (secret_bits[i] == '1')
            CX(q[i], q[n]);
    }
    H(q);
    Measure(q.head(n));
}

int main(int argc, char **argv) {
    std::string secret_bits(argv[1]);
    auto q = qalloc(secret_bits.size()+1);
    bernstein_vazirani(q, secret_bits);
    q.print();
}
\end{lstlisting}

Behind the scenes, the QCOR implementation uses the XACC
%(eXtreme-scale Accelerator)
framework to parse quantum assembly into quantum IR (Intermediate Representation, implemented as an n-ary tree of quantum gates), transform the IR (e.g., for optimizations), and communicate with accelerators~\cite{mccaskey_xacc_2020}. With its plugin-based architecture, adding an XACC plugin for a new accelerator exposes this new backend on the QCOR level; backends already exist for web APIs
%\aja{I originally said ``REST API'' but some readers may not know what a REST API is. is this too vague?}
for vendors such as IBM, IonQ, and Rigetti, as well as for calls to local simulator libraries. %However, XACC currently has no backends for local quantum hardware such as the GTRI testbed described in Section~\ref{sec:gtri-testbed}. %\ch{What does "local" mean here? Should be able to make a definitive statement, "is/is not" a backend as all relevant parties are authors!}

%% file: sections/03-impl.tex
\section{Backend Implementation}\label{sec:compiler-impl}
\subsection{Overview}
Our new backend consists of an XACC plugin with a new \texttt{Accelerator} implementation for the GTRI testbed. The backend can emit either quantum assembly or a sequence of primitive gates. Additionally, we add small modifications to the control software to read inputs from a file in a directory and write simulation outputs to the directory. Future work should evaluate a more robust queueing system than polling a directory on disk. We have released our backend code as open source online\footnote{
\ifdefined\doubleblind
    Link hidden for double-blind review
\else
    \url{https://github.com/ausbin/xacc/tree/ion-trap-backend}
\fi
}.

The remainder of this section describes how our backend prepares a quantum circuit for execution, focusing particularly on decomposing a circuit into primitive testbed operations. The decomposition consists of two passes: the first for two-qubit gates and the second for single-qubit gates, both detailed below. Each pass is an XACC \texttt{IRTransformation} which operates on XACC IR, replacing logical gates with either more logical gates or native operations. After the two passes, our backend converts the resulting XACC IR to a sequence (or table) of primitive operations and passes it to the control software. Our initial implementation for each compiler pass has time complexity $O(n^2)$, where $n$ is the number of program gates; linear-time implementations would be straightforward but we reserve them for future work, as they would make a negligible difference in runtime for our small benchmarks.

%\revtwo{If at all possible, it will be great if the authors can extend the gate reduction discussion, possibly studying the implication to the overall computation complexity.}\aja{Attempted to address the complexity of the compiler, which I'm guessing they mean here. If they mean the complexity of the resulting code, that doesn't make much sense to me}

\subsection{Two-Qubit Gate Compiler Pass}\label{sec:2q-pass}
The first decomposition pass decomposes two-qubit gates into a combination of the native two-qubit operation $XX(\pi/4)$ and logical single-qubit gates. When the first pass encounters a two-qubit gate in the XACC IR gateset other than a CNOT, such as a CZ or SWAP, the first pass decomposes it into CNOTs and single-qubit gates. Then, when the pass encounters a CNOT, including one it introduced, the pass decomposes the CNOT as shown in Fig.~\ref{fig:cnot-decomp}, leaving $XX(\pi/4)$ native operations as the only two-qubit gates in the IR.
%The Z-rotations shown in Fig.~\ref{fig:cnot-decomp} surrounding the MS gate would correct for the MS phase, as explained in Appendix A of~\cite{herold_universal_2016}, producing an effective $XX$ gate; however, we assume $\phi_L = \phi_R = 0$ and thus $MS = XX$ in this work, so we do not add these Z-rotations\ch{These extra left/right phases may add a lot of confusion. They may be an interesting degree of freedom to include in future work, but does it make sense to remove them here?}.

\subsection{Single-Qubit Gate Compiler Pass}\label{sec:1q-pass}
The second decomposition pass finds sequences of adjacent single-qubit gates operating on the same qubit, calculates the product $G$ of them, and uses a numerical optimizer to find the rotation angles for $R_\phi(\pi/2)$ native operations to approximate $G$ up to a user-configurable tolerance (default $10^{-4}$). This is similar to the existing compiler~\cite{herold_universal_2016}, but we have made some additional optimizations.

We use an L-BFGS optimizer to minimize an objective function that measures the ``distance'' between a sequence of rotations and the $2{\times}2$ goal unitary, starting with one rotation and adding rotations until the final objective function value is satisfactorily small. This approach, particularly the objective function definition, draws from how~\cite{lao_designing_2021}~decomposes $4{\times}4$ unitaries into native two-qubit gates.
%decompose two-qubit unitaries:

%More precisely,  $\Vec{\phi} = (\phi_1,\phi_2,\ldots,\phi_n)$ such that $A(\Vec{\phi}) \approx G$, where $A(\Vec{\phi})$ is defined in (\ref{eq:actual}).
With a $2{\times}2$ goal unitary $G$ and rotations $\Vec{\phi} = (\phi_1,\phi_2,\ldots,\phi_n)$, the optimization function to minimize is shown in (\ref{eq:obj-func2}).
Like the existing compiler \cite{herold_universal_2016}, we have always found $n \le 4$ in our testing; future work should prove that some gates require four $R_\phi(\pi/2)$ rotations, since a $Z$-$Y$ decomposition for example requires a maximum of only three rotations \cite{nielsen_quantum_2010}.
%Since only four rotations can produce any unitary \cite{maslov_basic_2017}\ch{Is this proven for arbitrary rotation $\theta$, or specifically for $\pi/2$?}, we know $n \le 4$.
The absolute value in (\ref{eq:obj-func1}) is squared to simplify finding the gradient $\nabla_{\Vec{\phi}}f$ from (\ref{eq:obj-func2}) using the product rule. We pre-computed this gradient for $n \in \{1,2,3,4\}$ using Mathematica and converted the result into parameterized C++ code.
\begin{align}
    A_\text{ex}(\Vec{\phi}) &= R_{\phi_n}(\pi/2)\cdots R_{\phi_2}(\pi/2)R_{\phi_1}(\pi/2) \label{eq:actual} \\
    f(\Vec{\phi}) &= 4 - \vert\tr(G^\dagger A(\Vec{\phi}))\vert^2 \label{eq:obj-func1} \\
                  &= 4 - \tr(G^\dagger A(\Vec{\phi}))\tr(G^\dagger A(\Vec{\phi}))\text{*} \label{eq:obj-func2}
\end{align}
%\ch{Does the `bar' above in (13) mean complex conjugate? How standard is that?}
%
Choosing $A_\text{ex}(\Vec{\phi})$ as the actual decomposition $A(\Vec{\phi})$ allows the aforementioned strategy to produce an approximately ``exact'' decomposition. We found we can make further optimizations by changing $A(\Vec{\phi})$ or skipping decomposition entirely depending on the situation. The following sections describe our situation-specific optimizations:
%Demonstrating the power of this numerical optimization--based approach, the following sections describe our optimizations on top of this approach:

\subsubsection{Decomposition up to an X Rotation} It is easy to show that $RX(\theta)$ commutes with $XX$, but $RY(\theta)$ and $RZ(\theta)$ only commute with $XX$ if $\theta \equiv 0 \bmod{2\pi}$ \cite{maslov_basic_2017}. (In this work, we use $RX$, $RY$, and $RZ$ as defined in \cite{nielsen_quantum_2010}.) Thus, if an $XX$ gate follows the sequence of single-qubit gates, we decompose $G$ up to $RX(\theta)$ for some $\theta$ and then commute the $RX(\theta)$ to the other side of the $XX$ gate, to be decomposed later. We achieve this with the optimizer by choosing $A_x(\Vec{\phi})$ as $A(\Vec{\phi})$ per the definition shown in (\ref{eq:actual_rx}). Note this requires a separate pre-computed gradient to pass to the optimizer.
\begin{align}
    A_x(\Vec{\phi}) &= RX(\phi_n)R_{\phi_{n-1}}(\pi/2)\cdots R_{\phi_2}(\pi/2)R_{\phi_1}(\pi/2) \label{eq:actual_rx}
\end{align}
In some rare cases ($0.05\%$ of cases we tested), the optimizer fails to find a decomposition up to an $RX(\theta)$ such that $f(\Vec{\phi})$ is less than the configured threshold, possibly due to floating point rounding errors. In such cases, we fall back to the exact decomposition $A_\text{ex}(\Vec{\phi})$ shown in (\ref{eq:actual}) to avoid harming fidelity.

\subsubsection{Decomposition up to a Z Rotation}\label{sec:rz-decomp} Relative phase shifts, i.e. $Z$ rotations, do not affect measurements when measuring in the standard computational basis of $Z$-eigenstates. So if a measurement follows the sequence of single-qubit gates, we decompose $G$ up to $RZ(\theta)$ for some $\theta$ and then discard the $RZ(\theta)$ gate entirely. Similar to the previous optimization, the numerical optimizer finds $\theta$ for us after we choose $A_z(\Vec{\phi})$ as $A(\Vec{\phi})$ per the definition below in (\ref{eq:actual_rz}). Note this again requires a separate pre-computed gradient to pass to the optimizer.
\begin{align}
    A_z(\Vec{\phi}) &= RZ(\phi_n)R_{\phi_{n-1}}(\pi/2)\cdots R_{\phi_2}(\pi/2)R_{\phi_1}(\pi/2) \label{eq:actual_rz}
\end{align}
The existing compiler has the ability to perform this optimization, which we have enabled in our evaluation.
\subsubsection{Decomposition starting with a Z Rotation} Immediately after state preparation, a qubit is in state $\ket{0}$, and $RZ(\theta)\ket{0} = \ket{0}$ up to a global phase regardless of angle $\theta$. Consequently, we ignore an initial $Z$ rotation when decomposing the first single-qubit gate sequence for any qubit not preceded by a two-qubit gate acting on that qubit. This optimization can be combined with the previous two optimizations as well as generation of an exact decomposition; we ask the numerical optimizer for from-Z-to-exact, from-Z-up-to-X, and from-Z-up-to-Z decompositions using the following respective definitions for $A(\Vec{\phi})$:
\begin{align}
    A_{z,\text{ex}}(\Vec{\phi}) &= R_{\phi_n}(\pi/2)\cdots R_{\phi_2}(\pi/2)RZ(\phi_1) \label{eq:actual_exact_from_rz} \\
    A_{z,x}(\Vec{\phi}) &= RX(\phi_n)R_{\phi_{n-1}}(\pi/2)\cdots R_{\phi_2}(\pi/2)RZ(\phi_1) \label{eq:actual_rx_from_rz} \\
    A_{z,z}(\Vec{\phi}) &= RZ(\phi_n)R_{\phi_{n-1}}(\pi/2)\cdots R_{\phi_2}(\pi/2)RZ(\phi_1) \label{eq:actual_rz_from_rz}
\end{align}
Similar to the previous optimization, we discard the $RZ(\phi_1)$ gate of the decomposition. Note that (\ref{eq:actual_exact_from_rz})-(\ref{eq:actual_rz_from_rz}) each require a new, separate pre-computed gradient for $f(\Vec{\phi})$.

The existing compiler supports the from-Z-to-exact case~\cite{herold_universal_2016}, producing a decomposition equivalent to (\ref{eq:actual_exact_from_rz}). We enable this optimization together with the previous exact-up-to-Z optimization in our evaluation of the existing compiler, henceforth calling this combination ``$RZ$ optimizations.''

\subsubsection{Ignoring Identity} If setting $A(\Vec{\phi})$ equal to the identity matrix $I$ and invoking $f(\Vec{\phi})$ reveals that $G$ is closer than the configured tolerance to $I$,
%\ch{What does global mean if $G$ is for one ion? Perhaps use `overall' instead?}
we skip generating any rotations at all.

\subsubsection{Discarding Trailing Gates}\label{sec:discard-trailing} If the sequence of gates immediately precedes the end of the circuit, without an explicit measurement by the programmer, we discard all the gates entirely, as they will not affect measurement outcomes.

%\ch{These are all sensible ideas. There's also a savings for gates following qubit initialization. Similar to 3), you can ignore an initial $RZ$ if you know you are in the $\ket{0}$ state.}
%To decompose single-qubit gates into $R_\phi(\pi/2)$ rotations, the second pass uses the same overall approach as the original GTRI compiler running in IGOR~\cite{herold_universal_2016}.
%The specific details resemble the approach \cite{lao_designing_2021}~uses to decompose two-qubit unitaries: we use L-BFGS to minimize an objective function that measures the `distance' between a sequence of rotations and the goal unitary, starting with one rotation and adding rotations until the final objective function value is satisfactorily small.
%
%\begin{align}
%    A(\Vec{\phi}) &= \prod_{i=n}^1 R_{\phi_i}(\pi/2) \\
%    f(\Vec{\phi}) &= 4 - \vert\tr(G^\dagger A(\Vec{\phi})\vert^2 \label{eq:obj-func1} \\
%                  &= 4 - \tr(G^\dagger A(\Vec{\phi})\overline{\tr(G^\dagger A(\Vec{\phi})} \label{eq:obj-func2}
%\end{align}

\subsection{Future Hardware Upgrades}\label{sec:future-opt}

\begin{table}[tbp]
\caption{Comparison of Native Operations Produced for the Bell State Circuit \texttt{H 0; CNOT 0,1} for Serial Single-Qubit Gates.}\label{tab:serial-ex}
\begin{center}
\begin{tabular}{lrr}
\hline
Operation & Target Ion & $\phi$ \\ \hline
$XX(\pi/4)$	& 0,1	& \\
$R\phi(\pi/2)$	& 0	& $\pi/2$ \\
$R\phi(\pi/2)$	& 1	& 0 \\ \hline
\end{tabular}
\end{center}
\end{table}

\begin{table}[tbp]
\caption{Comparison of Native Operations Produced for the Bell State Circuit \texttt{H 0; CNOT 0,1} for Parallel Single-Qubit Gates.}\label{tab:parallel-ex}
\begin{center}
\begin{tabular}{lrrrr}
\hline
Operation & Target Ion 1 & $\phi_1$ & Target Ion 2 & $\phi_2$ \\ \hline
$XX(\pi/4)$	& 0,1	& \\
$R\phi(\pi/2)$	& 0	& $\pi/2$ & 1	& 0 \\ \hline
\end{tabular}
\end{center}
\end{table}

In anticipation of future hardware upgrades, namely a tightly-focused beam for each individual ion, we implement rudimentary support for all-to-all qubit connectivity and parallel single-qubit operations in our compiler, either of which configuration flags can enable. %and the simulator included in the control software.

QCOR itself handles qubit placement, so we add support for full connectivity by simply having our backend pass a fully connected coupling map to QCOR instead of a coupling map indicating a linear chain. Full connectivity reduces the number of SWAP gates needed to execute a logical circuit on a linear chain of qubits, in turn reducing the number of CNOTs inserted by QCOR to perform SWAPs, and ultimately reducing the number of $XX(\pi/4)$ and $R_\phi(\pi/2)$ gates which together effect the swapping CNOTs.

We implement rudimentary support for parallel single-qubit using a na\"ive algorithm that greedily constructs a table from the XACC IR produced by the decomposition passes, with multiple $R_\phi(\pi/2)$ per row, and with each $XX(\pi/4)$ having its own row. Parallel single-qubit operations would not reduce the number of $XX(\pi/4)$ gates, but such a hardware upgrade could allow multiple $R_\phi(\pi/2)$ gates to occur across different qubits at once. Tables \ref{tab:serial-ex} and \ref{tab:parallel-ex} show examples of serial and parallel native operations, respectively, for the same input program. Clearly, we cannot compare runtime between serial and parallel configurations by counting the total number of gates; for example, Tables \ref{tab:serial-ex} and \ref{tab:parallel-ex} have the same number of $R_\phi(\pi/2)$ operations, but Table \ref{tab:parallel-ex} offers a shorter runtime. Thus, we estimate the length of the critical path by counting the number of rows in the table of native operations sent to the control software, henceforth calling each row a ``cycle'' whenever we need to distinguish from native gate counts.

Parallel two-qubit operations stand to reduce the number of cycles spent on $XX(\pi/4)$ gates\cite{figgatt_parallel_2019}. However, given parallel two-qubit gates cannot be implemented on our three-qubit benchmarks or the testbed in its original three-qubit configuration \cite{herold_universal_2016,fallek_transport_2016}. We leave investigating parallel two-qubit gates on the testbed as future work.
%\aja{Is this a cop-out?}\efd{no}\ch{agreed.}

%% file: sections/04-results.tex
\section{Evaluation}

\subsection{System Configuration}
Our targeted physical ion trap testbed and its control scheme have been modified for domain-specific computations based on global operations \cite{rajakumar_generating_2020}, so we cannot execute our benchmark circuits on the physical testbed itself. However, we can verify correct output of generated code by executing our operations in the simulator already included in the control software, originally used to aid in calibration. Rather than performing noisy simulations,
%which may be inaccurate or computationally expensive \aja{remove this clause?},
we instead choose to estimate the performance of our compilation by simply counting the number of primitive operations, and we argue why this is reasonable in Section \ref{sec:eval-gate-count}.
%We have no simulator that estimates noise, so we choose to estimate fidelity coarsely by counting primitive operations and argue why this is reasonable in Section \ref{sec:eval-gate-count}.

\subsection{Optimizations Tested}
We ran our benchmarks with the following three types of optimizations:
\begin{enumerate}
    \item High level optimizations already included in QCOR as described in \cite{nguyen_extending_2020}, which includes approaches such as~\cite{nam_automated_2018}~for simplifying large circuits
    \item Our single-qubit optimizations explained in Section \ref{sec:1q-pass}
    \item Our optimizations for future hardware mentioned in Section \ref{sec:future-opt}
\end{enumerate}
Given \#1 is designed for larger, more complex circuits on a higher number of qubits, we found \#1 made no difference in final native gate counts for our benchmarks. Consequently, for the remainder of this section we focus on the impact of \#2 and \#3. However, we note that less trivial quantum algorithms run on the testbed in the future could benefit significantly from the high-level QCOR optimizations, with \cite{nguyen_extending_2020} seeing an average 23.2\% reduction in gates on benchmarks ranging from 5 to 96 qubits~\cite{amy_staqfull-stack_2020}.
%\aja{Redescribe the optimizations tested briefly. Talk about QCOR optimizer and cite.}
%QCOR was applied but did not see benefit for the algorithms we tested.

\subsection{Benchmarks Chosen}
To evaluate our backend, we ran the following set of benchmark QCOR programs on three qubits:
\begin{itemize}
    \item GHZ (Greenberger-Horne-Zeilinger), which generates the state $\frac{1}{\sqrt{2}}\ket{000} + \frac{1}{\sqrt{2}}\ket{111}$
    \item Bernstein-Vazirani with secret string $s=11$, similar to Listing~\ref{lst:qcor-ex}
    \item Grover with one iteration and marked states $\ket{101}$ and $\ket{110}$ \cite{figgatt_complete_2017}
    \item Quantum Fourier Transform using the \texttt{qft()} routine included in QCOR
    \item VQE (Variational Quantum Eigensolver) on the three-qubit Hamiltonian from \cite{dumitrescu_cloud_2018} using the QCOR tooling for VQE
    %\item Quantum Fourier Transform on the state $\frac{1}{2}\left(\ket{001} + \ket{011} + \ket{101} + \ket{111}\right)$
    %\item Quantum Fourier Transform on the state $\frac{1}{2}\begin{bmatrix}0 & 1 & 0 & 1 & 0 & 1 & 0 & 1\end{bmatrix}$
\end{itemize}
%GHZ, QAOA, Hubbard (likely broken, maybe Deuteron instead?), Grover (maybe grouped with QAOA ``spiritually'')?

We ran each benchmark with the following three configurations of compiler and simulator:
\begin{enumerate}
    \item Against an existing XACC backend which runs the instructions from XACC IR directly on the local Quantum++ simulator~\cite{gheorghiu_quantum_2018}
    \item Against our new XACC backend configured to generate assembly, which the existing compiler compiles to a sequence of primitive operations for the control software to simulate
    \item Against our new XACC backend configured to generate a sequence of primitive operations on its own as described in Sections \ref{sec:2q-pass} through \ref{sec:future-opt}, also simulated by the control software
\end{enumerate}
To validate the results of our backend, we compare the probability distribution of measurements calculated from the final state vectors of \#1 through \#3. Next, to evaluate our optimizations for current hardware, we compare the number of native operations produced by the existing compiler in \#2 and our compiler in \#3. Finally, we evaluate our optimizations for possible future hardware by comparing the native gate count produced by \#2 with the native gate count produced by \#3 with future hardware optimizations enabled.
%\aja{Is this clear?}\ch{Yes, given the ``future hardware'' section above, I easily followed this.}

%In both the latter cases, IGOR simulates the sequence of primitive operations and returns measurement results for 8192 runs.
%\ch{What is the Quantum++ simulator, or the Igor simulation for? Presumably, you're comparing that they give the same output? If so, I'd be clear about that.}

% \aja{removed this paragraph as it simply does not matter}
%For VQE, we calculated the expectation of $Z$, $\langle \psi \vert Z \vert \psi \rangle$, directly from the final state $\ket{\psi}$ calculated by the IGOR simulator rather than from simulated measurement counts. \ch{Is Quantum++ not a state-vector simulator as well? It makes a ton of sense to take measurement out of the comparison and look at the output state vector (since measurement will just sample this).}Future work should determine the proper variational optimizer configuration and the satisfactory number of simulator runs such that one can calculate $\langle \psi \vert Z \vert \psi \rangle$ strictly from measurement results without the variational optimizer struggling to converge.

\subsection{Validation Results}
%\begin{table}[htbp]
%\begin{table}
%\caption{Benchmark Primitive Operation Counts}
%\begin{center}
%\begin{tabular}{c|cc|cc}
%\hline
%Benchmark & Old MS & New MS & Old $R_\phi(\pi/2)$ & New $R_\phi(\pi/2)$ \\
%          & count & count & count & count \\ \hline
%BV & 8 & 8 & 50 & 23 \\
%GHZ & 2 & 2 & 11 & 7 \\
%Grover & 30 & 30 & 185 & 100 \\
%QFT & 18 & 18 & 107 & 68 \\
%VQE$^\dagger$ & 10 & 10 & 55.68 & 33.68 \\
%\hline
%\multicolumn{5}{l}{$^\dagger$Averaged across all ansatz executions}
%\end{tabular}
%\label{tab:opcounts}
%\end{center}
%\end{table}

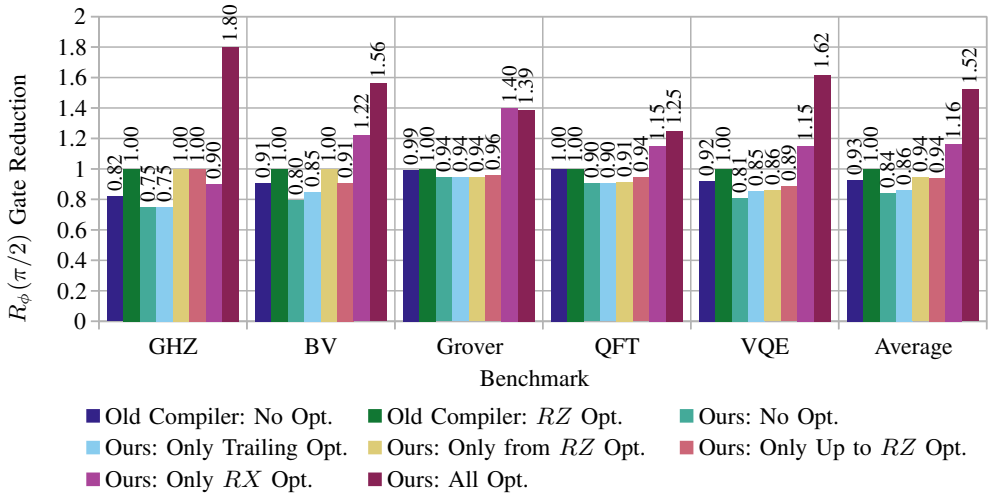
\begin{figure*}[th]
    %\begin{center}{\def\svgwidth{0.75\linewidth}\small\input{fig/pirot-speedup}}\end{center}%
    \centering{\def\svgwidth{0.75\linewidth}\small\input{fig/pirot-speedup-text}}%
    \caption{Comparison of reduction in $R_\phi(\pi/2)$ operations generated across our benchmarks by the existing compiler and by other compiler configurations (higher is better). We calculate reduction by dividing the $R_\phi(\pi/2)$ gates the existing compiler generates with $RZ$ optimizations enabled by the number of $R_\phi(\pi/2)$ gates generated by some other configuration. We include the existing compiler without optimizations for comparison.}\label{fig:pirot-speedup-graph}
\end{figure*}

To validate the results of our benchmarks, we calculated the probability distribution of measurements from the final state vector produced by the Quantum++ simulator and control system simulation of the sequence of primitive operations generated by both the existing compiler and ours. The VQE benchmark represents a special case in that instead of simply running a quantum kernel, the ansatz quantum kernel acts as part of the objective function for an optimizer on the variational parameters. As a result, in addition to comparing the probability of different measurements of the first VQE iteration (subsequent iterations diverged in parameters), we ensured the VQE benchmark converged to a result approximately equal to the value found in \cite{dumitrescu_cloud_2018}. 

The resulting states from our compiler often did not match those produced by either Quantum++ or the control software simulation of the existing compiler results, even up to a global phase, but this is intended behavior of the single-qubit pass described in Section \ref{sec:1q-pass}. First, the optimization in Section \ref{sec:rz-decomp} discards trailing Z-rotations, introducing relative phase differences in the final state. Second, the discarding of trailing gates mentioned in Section \ref{sec:discard-trailing} introduces some differences in final state compared to the others; for example, with Bernstein-Vazirani, our compiler produces final state $\frac{1}{\sqrt{2}}\ket{110} - \frac{1}{\sqrt{2}}\ket{111}$ rather than $\ket{111}$ owing to an elided final Hadamard gate on the ancilla qubit, which our program does not explicitly \texttt{Measure}\footnote{On real hardware, one might choose to measure the ancilla to verify that it is \ket{1} as an error detection measure, but we leave the BV benchmark as-is to ensure we test how our compiler handles gates on unmeasured qubits.}.
%\ch{May want to call qubit 2 the `ancilla' to be clearer. While this ancilla doesn't need to be measured to read out hidden string, it should be \ket{1} at the end of the algorithm and can herald an error if it's not; so in practice it's typically measured. I don't think you need to change your compiler.}
In all cases, the differences in final quantum state do not affect the probability distribution of measurements.
%\ch{So perhaps you should be comparing $\mathrm{Pr}(\mbox{observe each bit string})$, computed from the state vector instead of inferred through a finite number of measurements?}

%This is equivalent to comparing final state vectors up to relative and global phase; the former matters because of 

%\textbf{TODO:} How do we validate the output is correct? 1) Verify correct execution and 2) check state from the Igor Pro simulation backend. Faster simulation time (decreasing gate count, capability to run more complicated programs)?

%\aja{TODO: Is there any difference in simulation time with reductions in gate count?} 
%\ch{For such small circuits, I don't imagine this is important. For large enough circuits, full state-vector simulation wont' be possible! Maybe first ask yourself what you are using the simulator for here before discussing this?}

\subsection{Gate Count Comparison}\label{sec:eval-gate-count}
\begin{figure}
    {\def\svgwidth{\linewidth}\small\input{fig/pirot-counts-text}}
\caption{Comparison of $R_\phi(\pi/2)$ operation counts generated across our benchmarks by the existing compiler and by different configurations of our compiler (lower is better).}\label{fig:pirot-count-graph}
\end{figure}
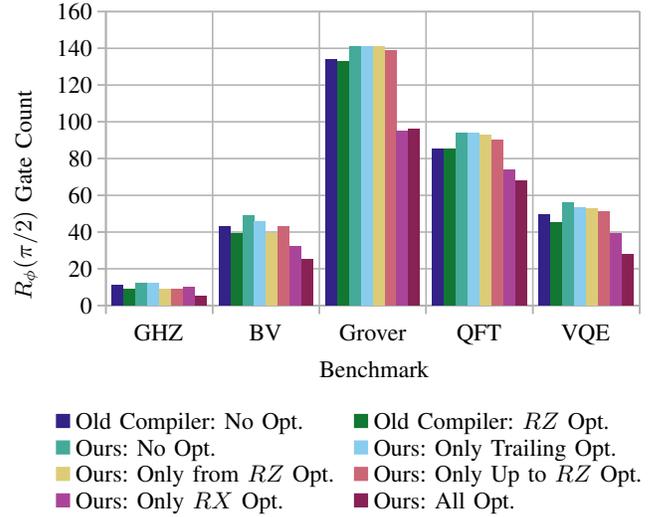

%\begin{table}
%\caption{Benchmark Primitive Operation Counts}
%\begin{center}
%\begin{tabular}{c|c}
%\hline
%Optimization & Average \\ \hline
%
%VQE$^\dagger$ & 10 & 10 & 55.68 & 33.68 \\
%\hline
%\multicolumn{5}{l}{$^\dagger$Averaged across all ansatz executions}
%\end{tabular}
%\label{tab:opcounts}
%\end{center}
%\end{table}

We find that as expected, the two-qubit pass described in Section \ref{sec:2q-pass} eliminates no $XX(\pi/4)$ gates, leaving our count of $XX(\pi/4)$ gates identical to the previous compiler. However, our one-qubit pass in Section \ref{sec:1q-pass} achieves an average $1.54\times$ reduction in $R_\phi(\pi/2)$ operations, even with the $RZ$ optimizations enabled in the existing compiler.
Fig.~\ref{fig:pirot-count-graph} shows a comparison of primitive operation counts between our compiler and the existing compiler for all our benchmarks.

To examine which optimizations contributed most to the reduction in single-qubit operations, we also compiled the benchmarks with only individual optimizations from Section \ref{sec:1q-pass} enabled, and also none of them enabled. The $R_\phi(\pi/2)$ count reduction numbers plotted in Fig.~\ref{fig:pirot-speedup-graph} show that individual optimizations struggle to compete with the $RZ$ optimizations in the existing compiler, with the exception of our $RX$ optimizations usually offering some individual reduction. On the GHZ benchmark, however, the $RX$ freedom alone is not as effective, with all optimizations combined performing better. Indeed, on average, each optimization provides a modest reduction in gates, but the combination readily beats any individual optimization.
%. gates on unmeasured qubits and $Z$-rotations at the end show modest improvements (an average of $1.10\times$ and $1.20\times$ on their own, respectively), commuting $X$-rotations around $XX$ gates deliver the bulk of gate reductions, at an average of $1.50\times$ on its own. \ch{The GHZ case stands out as an outlier; the $R_x$ freedom alone is not very effective, yet everything combined does much better.}

The error and duration estimates in \cite{maslov_basic_2017} for $R_\phi(\theta)$ gates on ion trap systems is defined in terms of $\theta$, not $\phi$, so with $\theta$ fixed to $\pi/2$, the number of $R_\phi(\pi/2)$ gates reasonably correlates with error and duration. Consequently, we expect that our benchmarks and other quantum circuits with similar complexity will experience higher overall fidelity on the testbed using our compiler. Furthermore, an $n\times$ reduction in physical $R_\phi(\pi/2)$ gates could allow $n\times$ more sequences of adjacent logical single-qubit gates with the same approximate total duration and total error as before. Thus, if combined with future optimizations for two-qubit gates%
%\footnote{We cannot generate physical rotations corresponding to sequences of logical single-qubit gates without two-qubit gates separating groups of single-qubit gates.}
, our compiler could allow running longer, more complicated programs on the testbed.

\subsection{Impact of Hardware Upgrades}
Our compiler does not reduce the number of $XX(\pi/4)$ gates on present hardware, but we found fully connected qubits can reduce the $XX(\pi/4)$ gate count by $2.40{\times}$ on average, as seen in Fig.~\ref{fig:ms-hw-speedup-graph}. Our Bernstein-Vazirani benchmark showed the greatest reduction at $4.00{\times}$, going from 8 operations to 2 operations, owing to the removal of six CNOTs used for two SWAPs. GHZ, on the other hand, showed no benefit, as it executes CNOTs only on adjacent ions. In general, then, we see that the benefit of full connectivity is proportional to the share of of two-qubit gates between non-adjacent ions.

\begin{figure}
    {\def\svgwidth{\linewidth}\small\input{fig/pirot-hw-speedup-text}}%
    %\centering{\def\svgwidth{0.75\linewidth}\small\input{fig/pirot-hw-speedup}}%
\caption{Comparison of reduction in $R_\phi(\pi/2)$ operations generated across our benchmarks by our compiler between present hardware and different hardware upgrades (higher is better).}\label{fig:pirot-hw-speedup-graph}
\end{figure}
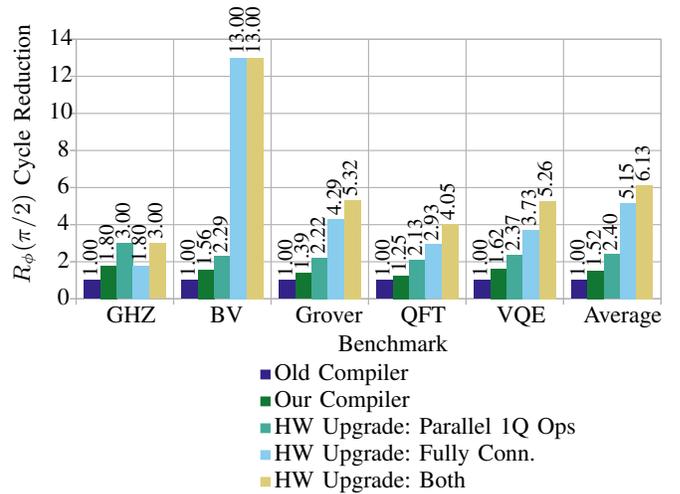

As shown in Fig.~\ref{fig:pirot-hw-speedup-graph}, full connectivity can also decrease the number of $R_\phi(\pi/2)$ operations since we use them to effect a CNOT using $XX(\pi/4)$ (see Fig.~\ref{fig:cnot-decomp}). With a $5.13{\times}$ reduction on average, a fully-connected upgrade could readily beat the average $1.52{\times}$ reduction seen with our compiler on current hardware. Still, some benchmarks see no benefit, such as GHZ for the reasons previously mentioned.

Although it cannot reduce $XX(\pi/4)$ cycles, Fig.~\ref{fig:pirot-hw-speedup-graph} shows that hardware support for parallel $R_\phi(\pi/2)$ operations on different qubits could reduce $R_\phi(\pi/2)$ cycles by an average of $2.40{\times}$ over the existing compiler and hardware regime, compared to $1.52{\times}$ with our compiler on present hardware. While all benchmarks saw at least some parallelism, future work could likely increase this reduction by investigating more complex strategies for single-qubit unitary decomposition (Section \ref{sec:1q-pass}) that take parallelism into account.

%\aja{What is the citation we are basing these HW optimizations on - forward looking}

\begin{figure}
    {\def\svgwidth{\linewidth}\small\input{fig/ms-hw-speedup-text}}
\caption{Comparison of reduction in $XX(\pi/4)$ operation counts generated across our benchmarks by our compiler between present hardware and upgraded hardware with all-to-all qubit connectivity (higher is better).}\label{fig:ms-hw-speedup-graph}
\end{figure}
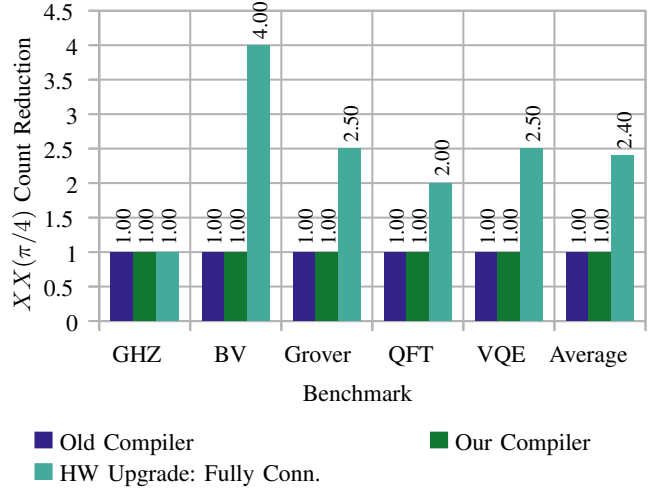

%% file: fig/pirot-speedup-text.tex
%% Creator: Inkscape 1.1.1 (1:1.1+202109281943+c3084ef5ed), www.inkscape.org
%% Accompanies image file 'pirot-speedup.pdf' (pdf, eps, ps)
\begingroup%
  \makeatletter%
  \providecommand\color[2][]{%
    \errmessage{(Inkscape) Color is used for the text in Inkscape, but the package 'color.sty' is not loaded}%
    \renewcommand\color[2][]{}%
  }%
  \providecommand\transparent[1]{%
    \errmessage{(Inkscape) Transparency is used (non-zero) for the text in Inkscape, but the package 'transparent.sty' is not loaded}%
    \renewcommand\transparent[1]{}%
  }%
  \providecommand\rotatebox[2]{#2}%
  \newcommand*\fsize{\dimexpr\f@size pt\relax}%
  \newcommand*\lineheight[1]{\fontsize{\fsize}{#1\fsize}\selectfont}%
  \ifx\svgwidth\undefined%
    \setlength{\unitlength}{540.9070866bp}%
    \ifx\svgscale\undefined%
      \relax%
    \else%
      \setlength{\unitlength}{\unitlength * \real{\svgscale}}%
    \fi%
  \else%
    \setlength{\unitlength}{\svgwidth}%
  \fi%
  \global\let\svgwidth\undefined%
  \global\let\svgscale\undefined%
  \makeatother%
  \begin{picture}(1,0.49506023)%
    \lineheight{1}%
    \setlength\tabcolsep{0pt}%
    \put(0,0){\includegraphics[width=\unitlength,page=1]{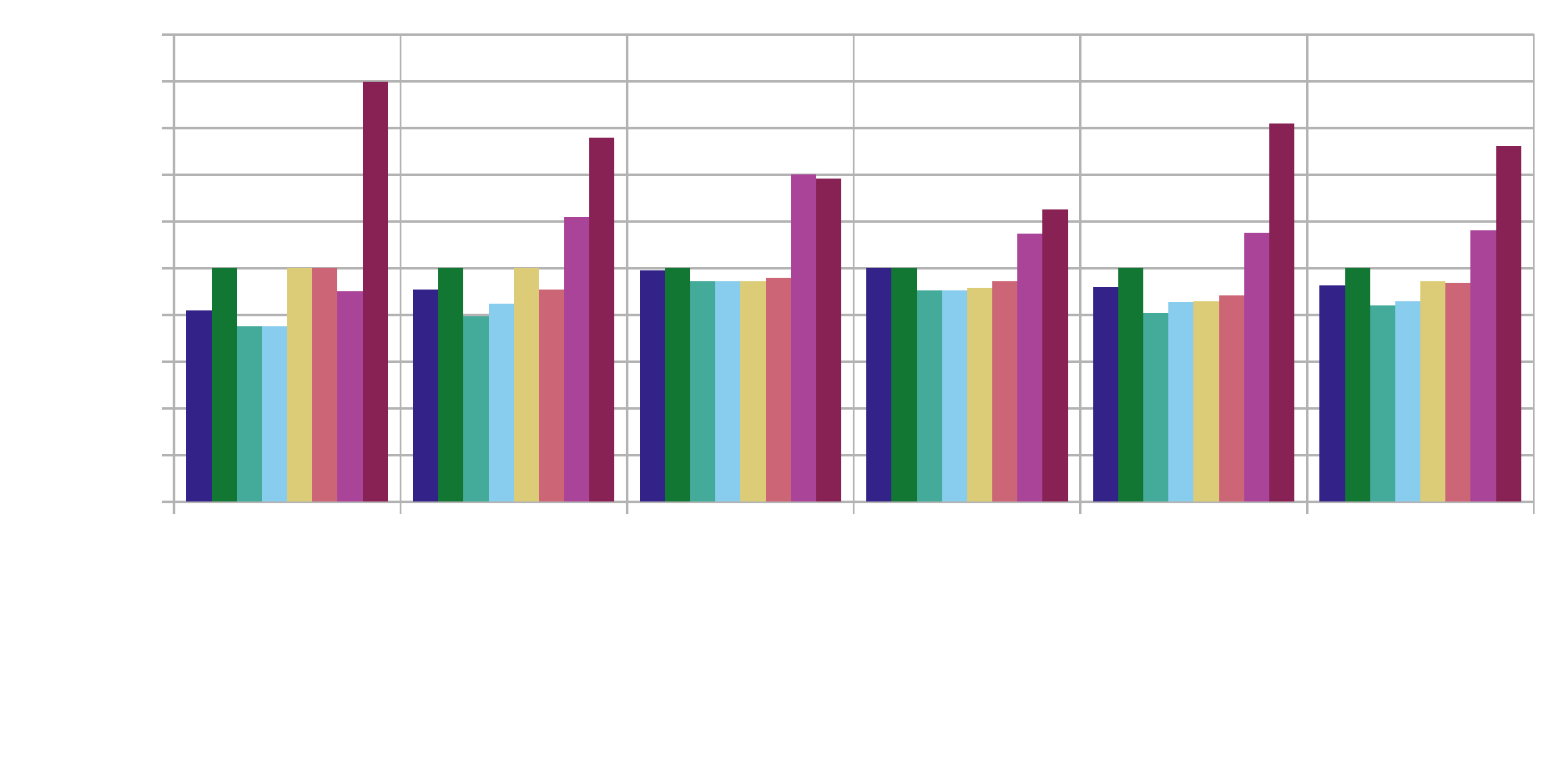}}%
    \put(0.15962687,0.1424903){\makebox(0,0)[lt]{\lineheight{1.25}\smash{\begin{tabular}[t]{l}GHZ\end{tabular}}}}%
    \put(0.31186458,0.1424903){\makebox(0,0)[lt]{\lineheight{1.25}\smash{\begin{tabular}[t]{l}BV\end{tabular}}}}%
    \put(0.44120113,0.1424903){\makebox(0,0)[lt]{\lineheight{1.25}\smash{\begin{tabular}[t]{l}Grover\end{tabular}}}}%
    \put(0.59558746,0.1424903){\makebox(0,0)[lt]{\lineheight{1.25}\smash{\begin{tabular}[t]{l}QFT\end{tabular}}}}%
    \put(0.73807777,0.1424903){\makebox(0,0)[lt]{\lineheight{1.25}\smash{\begin{tabular}[t]{l}VQE\end{tabular}}}}%
    \put(0.86882926,0.1424903){\makebox(0,0)[lt]{\lineheight{1.25}\smash{\begin{tabular}[t]{l}Average\end{tabular}}}}%
    \put(0.08673095,0.16727806){\makebox(0,0)[lt]{\lineheight{1.25}\smash{\begin{tabular}[t]{l}0\end{tabular}}}}%
    \put(0.07011844,0.19714915){\makebox(0,0)[lt]{\lineheight{1.25}\smash{\begin{tabular}[t]{l}0.2\end{tabular}}}}%
    \put(0.07011844,0.22691542){\makebox(0,0)[lt]{\lineheight{1.25}\smash{\begin{tabular}[t]{l}0.4\end{tabular}}}}%
    \put(0.07011844,0.25673409){\makebox(0,0)[lt]{\lineheight{1.25}\smash{\begin{tabular}[t]{l}0.6\end{tabular}}}}%
    \put(0.07011844,0.28655277){\makebox(0,0)[lt]{\lineheight{1.25}\smash{\begin{tabular}[t]{l}0.8\end{tabular}}}}%
    \put(0.08673095,0.31631904){\makebox(0,0)[lt]{\lineheight{1.25}\smash{\begin{tabular}[t]{l}1\end{tabular}}}}%
    \put(0.07011844,0.34613772){\makebox(0,0)[lt]{\lineheight{1.25}\smash{\begin{tabular}[t]{l}1.2\end{tabular}}}}%
    \put(0.07011844,0.3759564){\makebox(0,0)[lt]{\lineheight{1.25}\smash{\begin{tabular}[t]{l}1.4\end{tabular}}}}%
    \put(0.07011844,0.40572267){\makebox(0,0)[lt]{\lineheight{1.25}\smash{\begin{tabular}[t]{l}1.6\end{tabular}}}}%
    \put(0.07011844,0.43554135){\makebox(0,0)[lt]{\lineheight{1.25}\smash{\begin{tabular}[t]{l}1.8\end{tabular}}}}%
    \put(0.08673095,0.46530762){\makebox(0,0)[lt]{\lineheight{1.25}\smash{\begin{tabular}[t]{l}2\end{tabular}}}}%
    \put(0.13331936,0.30216958){\rotatebox{90}{\makebox(0,0)[lt]{\lineheight{1.25}\smash{\begin{tabular}[t]{l}\footnotesize 0.82\end{tabular}}}}}%
    \put(0.27795829,0.31537575){\rotatebox{90}{\makebox(0,0)[lt]{\lineheight{1.25}\smash{\begin{tabular}[t]{l}\footnotesize 0.91\end{tabular}}}}}%
    \put(0.42254481,0.32816267){\rotatebox{90}{\makebox(0,0)[lt]{\lineheight{1.25}\smash{\begin{tabular}[t]{l}\footnotesize 0.99\end{tabular}}}}}%
    \put(0.56713133,0.32926318){\rotatebox{90}{\makebox(0,0)[lt]{\lineheight{1.25}\smash{\begin{tabular}[t]{l}\footnotesize 1.00\end{tabular}}}}}%
    \put(0.71171785,0.31715753){\rotatebox{90}{\makebox(0,0)[lt]{\lineheight{1.25}\smash{\begin{tabular}[t]{l}\footnotesize 0.92\end{tabular}}}}}%
    \put(0.85635678,0.31841526){\rotatebox{90}{\makebox(0,0)[lt]{\lineheight{1.25}\smash{\begin{tabular}[t]{l}\footnotesize 0.93\end{tabular}}}}}%
    \put(0.14940782,0.32926318){\rotatebox{90}{\makebox(0,0)[lt]{\lineheight{1.25}\smash{\begin{tabular}[t]{l}\footnotesize 1.00\end{tabular}}}}}%
    \put(0.29399434,0.32926318){\rotatebox{90}{\makebox(0,0)[lt]{\lineheight{1.25}\smash{\begin{tabular}[t]{l}\footnotesize 1.00\end{tabular}}}}}%
    \put(0.43863327,0.32926318){\rotatebox{90}{\makebox(0,0)[lt]{\lineheight{1.25}\smash{\begin{tabular}[t]{l}\footnotesize 1.00\end{tabular}}}}}%
    \put(0.58316738,0.32926318){\rotatebox{90}{\makebox(0,0)[lt]{\lineheight{1.25}\smash{\begin{tabular}[t]{l}\footnotesize 1.00\end{tabular}}}}}%
    \put(0.72780631,0.32926318){\rotatebox{90}{\makebox(0,0)[lt]{\lineheight{1.25}\smash{\begin{tabular}[t]{l}\footnotesize 1.00\end{tabular}}}}}%
    \put(0.87239283,0.32926318){\rotatebox{90}{\makebox(0,0)[lt]{\lineheight{1.25}\smash{\begin{tabular}[t]{l}\footnotesize 1.00\end{tabular}}}}}%
    \put(0.16544387,0.29200293){\rotatebox{90}{\makebox(0,0)[lt]{\lineheight{1.25}\smash{\begin{tabular}[t]{l}\footnotesize 0.75\end{tabular}}}}}%
    \put(0.3100828,0.29881564){\rotatebox{90}{\makebox(0,0)[lt]{\lineheight{1.25}\smash{\begin{tabular}[t]{l}\footnotesize 0.80\end{tabular}}}}}%
    \put(0.45466932,0.3207735){\rotatebox{90}{\makebox(0,0)[lt]{\lineheight{1.25}\smash{\begin{tabular}[t]{l}\footnotesize 0.94\end{tabular}}}}}%
    \put(0.59925584,0.3149565){\rotatebox{90}{\makebox(0,0)[lt]{\lineheight{1.25}\smash{\begin{tabular}[t]{l}\footnotesize 0.90\end{tabular}}}}}%
    \put(0.74384236,0.30080704){\rotatebox{90}{\makebox(0,0)[lt]{\lineheight{1.25}\smash{\begin{tabular}[t]{l}\footnotesize 0.81\end{tabular}}}}}%
    \put(0.88848129,0.30547112){\rotatebox{90}{\makebox(0,0)[lt]{\lineheight{1.25}\smash{\begin{tabular}[t]{l}\footnotesize 0.84\end{tabular}}}}}%
    \put(0.18153233,0.29200293){\rotatebox{90}{\makebox(0,0)[lt]{\lineheight{1.25}\smash{\begin{tabular}[t]{l}\footnotesize 0.75\end{tabular}}}}}%
    \put(0.32611886,0.30657164){\rotatebox{90}{\makebox(0,0)[lt]{\lineheight{1.25}\smash{\begin{tabular}[t]{l}\footnotesize 0.85\end{tabular}}}}}%
    \put(0.47075778,0.3207735){\rotatebox{90}{\makebox(0,0)[lt]{\lineheight{1.25}\smash{\begin{tabular}[t]{l}\footnotesize 0.94\end{tabular}}}}}%
    \put(0.6152919,0.3149565){\rotatebox{90}{\makebox(0,0)[lt]{\lineheight{1.25}\smash{\begin{tabular}[t]{l}\footnotesize 0.90\end{tabular}}}}}%
    \put(0.75993082,0.30751494){\rotatebox{90}{\makebox(0,0)[lt]{\lineheight{1.25}\smash{\begin{tabular}[t]{l}\footnotesize 0.85\end{tabular}}}}}%
    \put(0.90451735,0.30840583){\rotatebox{90}{\makebox(0,0)[lt]{\lineheight{1.25}\smash{\begin{tabular}[t]{l}\footnotesize 0.86\end{tabular}}}}}%
    \put(0.19762079,0.32926318){\rotatebox{90}{\makebox(0,0)[lt]{\lineheight{1.25}\smash{\begin{tabular}[t]{l}\footnotesize 1.00\end{tabular}}}}}%
    \put(0.34220732,0.32926318){\rotatebox{90}{\makebox(0,0)[lt]{\lineheight{1.25}\smash{\begin{tabular}[t]{l}\footnotesize 1.00\end{tabular}}}}}%
    \put(0.48684624,0.3207735){\rotatebox{90}{\makebox(0,0)[lt]{\lineheight{1.25}\smash{\begin{tabular}[t]{l}\footnotesize 0.94\end{tabular}}}}}%
    \put(0.63138036,0.31642385){\rotatebox{90}{\makebox(0,0)[lt]{\lineheight{1.25}\smash{\begin{tabular}[t]{l}\footnotesize 0.91\end{tabular}}}}}%
    \put(0.77601929,0.30830102){\rotatebox{90}{\makebox(0,0)[lt]{\lineheight{1.25}\smash{\begin{tabular}[t]{l}\footnotesize 0.86\end{tabular}}}}}%
    \put(0.92060581,0.3207735){\rotatebox{90}{\makebox(0,0)[lt]{\lineheight{1.25}\smash{\begin{tabular}[t]{l}\footnotesize 0.94\end{tabular}}}}}%
    \put(0.21365685,0.32926318){\rotatebox{90}{\makebox(0,0)[lt]{\lineheight{1.25}\smash{\begin{tabular}[t]{l}\footnotesize 1.00\end{tabular}}}}}%
    \put(0.35829578,0.31537575){\rotatebox{90}{\makebox(0,0)[lt]{\lineheight{1.25}\smash{\begin{tabular}[t]{l}\footnotesize 0.91\end{tabular}}}}}%
    \put(0.50282989,0.32281731){\rotatebox{90}{\makebox(0,0)[lt]{\lineheight{1.25}\smash{\begin{tabular}[t]{l}\footnotesize 0.96\end{tabular}}}}}%
    \put(0.64746882,0.32098313){\rotatebox{90}{\makebox(0,0)[lt]{\lineheight{1.25}\smash{\begin{tabular}[t]{l}\footnotesize 0.94\end{tabular}}}}}%
    \put(0.79205534,0.31212661){\rotatebox{90}{\makebox(0,0)[lt]{\lineheight{1.25}\smash{\begin{tabular}[t]{l}\footnotesize 0.89\end{tabular}}}}}%
    \put(0.93669427,0.32009223){\rotatebox{90}{\makebox(0,0)[lt]{\lineheight{1.25}\smash{\begin{tabular}[t]{l}\footnotesize 0.94\end{tabular}}}}}%
    \put(0.22974531,0.31432764){\rotatebox{90}{\makebox(0,0)[lt]{\lineheight{1.25}\smash{\begin{tabular}[t]{l}\footnotesize 0.90\end{tabular}}}}}%
    \put(0.37433183,0.36185934){\rotatebox{90}{\makebox(0,0)[lt]{\lineheight{1.25}\smash{\begin{tabular}[t]{l}\footnotesize 1.22\end{tabular}}}}}%
    \put(0.51891835,0.38884813){\rotatebox{90}{\makebox(0,0)[lt]{\lineheight{1.25}\smash{\begin{tabular}[t]{l}\footnotesize 1.40\end{tabular}}}}}%
    \put(0.66350487,0.35137826){\rotatebox{90}{\makebox(0,0)[lt]{\lineheight{1.25}\smash{\begin{tabular}[t]{l}\footnotesize 1.15\end{tabular}}}}}%
    \put(0.8081438,0.35158788){\rotatebox{90}{\makebox(0,0)[lt]{\lineheight{1.25}\smash{\begin{tabular}[t]{l}\footnotesize 1.15\end{tabular}}}}}%
    \put(0.95273032,0.35357929){\rotatebox{90}{\makebox(0,0)[lt]{\lineheight{1.25}\smash{\begin{tabular}[t]{l}\footnotesize 1.16\end{tabular}}}}}%
    \put(0.24578136,0.44843308){\rotatebox{90}{\makebox(0,0)[lt]{\lineheight{1.25}\smash{\begin{tabular}[t]{l}\footnotesize 1.80\end{tabular}}}}}%
    \put(0.39042029,0.41269259){\rotatebox{90}{\makebox(0,0)[lt]{\lineheight{1.25}\smash{\begin{tabular}[t]{l}\footnotesize 1.56\end{tabular}}}}}%
    \put(0.53495441,0.3866471){\rotatebox{90}{\makebox(0,0)[lt]{\lineheight{1.25}\smash{\begin{tabular}[t]{l}\footnotesize 1.39\end{tabular}}}}}%
    \put(0.67959333,0.36652343){\rotatebox{90}{\makebox(0,0)[lt]{\lineheight{1.25}\smash{\begin{tabular}[t]{l}\footnotesize 1.25\end{tabular}}}}}%
    \put(0.82417986,0.42139189){\rotatebox{90}{\makebox(0,0)[lt]{\lineheight{1.25}\smash{\begin{tabular}[t]{l}\footnotesize 1.62\end{tabular}}}}}%
    \put(0.96881878,0.40708521){\rotatebox{90}{\makebox(0,0)[lt]{\lineheight{1.25}\smash{\begin{tabular}[t]{l}\footnotesize 1.52\end{tabular}}}}}%
    \put(0,0){\includegraphics[width=\unitlength,page=2]{pirot-speedup.pdf}}%
    \put(0.11712609,0.07483492){\makebox(0,0)[lt]{\lineheight{1.25}\smash{\begin{tabular}[t]{l}Old Compiler: No Opt.\end{tabular}}}}%
    \put(0.39262132,0.07483492){\makebox(0,0)[lt]{\lineheight{1.25}\smash{\begin{tabular}[t]{l}Old Compiler: $RZ$ Opt.\end{tabular}}}}%
    \put(0.69720155,0.07483492){\makebox(0,0)[lt]{\lineheight{1.25}\smash{\begin{tabular}[t]{l}Ours: No Opt.\end{tabular}}}}%
    \put(0.11712609,0.04381092){\makebox(0,0)[lt]{\lineheight{1.25}\smash{\begin{tabular}[t]{l}Ours: Only Trailing Opt.\end{tabular}}}}%
    \put(0.39262132,0.04381092){\makebox(0,0)[lt]{\lineheight{1.25}\smash{\begin{tabular}[t]{l}Ours: Only from $RZ$ Opt.\end{tabular}}}}%
    \put(0.69720155,0.04381092){\makebox(0,0)[lt]{\lineheight{1.25}\smash{\begin{tabular}[t]{l}Ours: Only Up to $RZ$ Opt.\end{tabular}}}}%
    \put(0.11712609,0.01273451){\makebox(0,0)[lt]{\lineheight{1.25}\smash{\begin{tabular}[t]{l}Ours: Only $RX$ Opt.\end{tabular}}}}%
    \put(0.39262132,0.01273451){\makebox(0,0)[lt]{\lineheight{1.25}\smash{\begin{tabular}[t]{l}Ours: All Opt.\end{tabular}}}}%
    \put(0.48323027,0.11193795){\makebox(0,0)[lt]{\lineheight{1.25}\smash{\begin{tabular}[t]{l}Benchmark\end{tabular}}}}%
    \put(0.03925165,0.17325228){\rotatebox{90}{\makebox(0,0)[lt]{\lineheight{1.25}\smash{\begin{tabular}[t]{l}$R_\phi(\pi/2)$ Gate Reduction\end{tabular}}}}}%
  \end{picture}%
\endgroup%

%% file: fig/pirot-counts-text.tex
%% Creator: Inkscape 1.1 (1:1.1+202106031931+af4d65493e), www.inkscape.org
%% Accompanies image file 'pirot-counts.pdf' (pdf, eps, ps)
\begingroup%
  \makeatletter%
  \providecommand\color[2][]{%
    \errmessage{(Inkscape) Color is used for the text in Inkscape, but the package 'color.sty' is not loaded}%
    \renewcommand\color[2][]{}%
  }%
  \providecommand\transparent[1]{%
    \errmessage{(Inkscape) Transparency is used (non-zero) for the text in Inkscape, but the package 'transparent.sty' is not loaded}%
    \renewcommand\transparent[1]{}%
  }%
  \providecommand\rotatebox[2]{#2}%
  \newcommand*\fsize{\dimexpr\f@size pt\relax}%
  \newcommand*\lineheight[1]{\fontsize{\fsize}{#1\fsize}\selectfont}%
  \ifx\svgwidth\undefined%
    \setlength{\unitlength}{338.71180845bp}%
    \ifx\svgscale\undefined%
      \relax%
    \else%
      \setlength{\unitlength}{\unitlength * \real{\svgscale}}%
    \fi%
  \else%
    \setlength{\unitlength}{\svgwidth}%
  \fi%
  \global\let\svgwidth\undefined%
  \global\let\svgscale\undefined%
  \makeatother%
  \begin{picture}(1,0.79136329)%
    \lineheight{1}%
    \setlength\tabcolsep{0pt}%
    \put(0,0){\includegraphics[width=\unitlength,page=1]{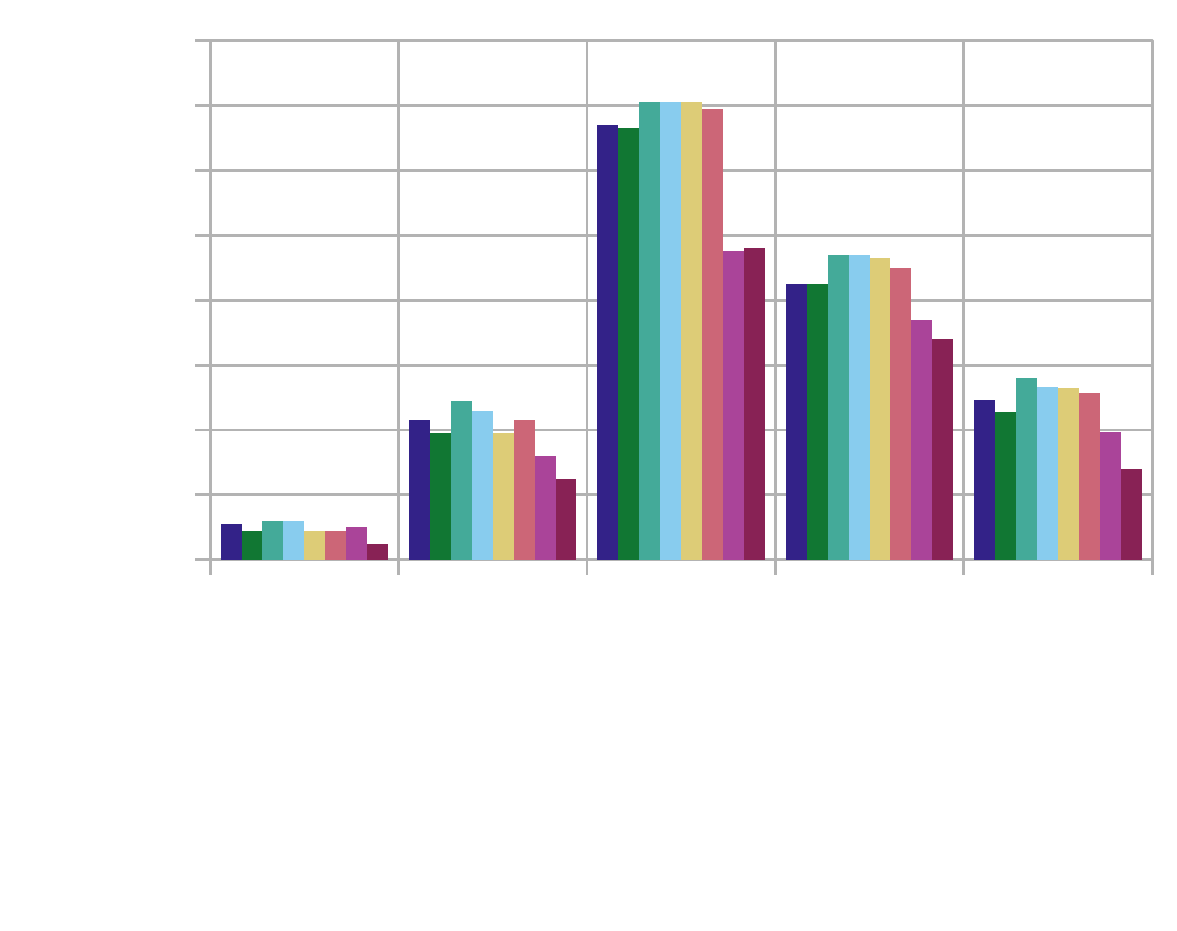}}%
    \put(0.22110637,0.26345301){\makebox(0,0)[lt]{\lineheight{1.25}\smash{\begin{tabular}[t]{l}GHZ\end{tabular}}}}%
    \put(0.39342204,0.26345301){\makebox(0,0)[lt]{\lineheight{1.25}\smash{\begin{tabular}[t]{l}BV\end{tabular}}}}%
    \put(0.529333,0.26345301){\makebox(0,0)[lt]{\lineheight{1.25}\smash{\begin{tabular}[t]{l}Grover\end{tabular}}}}%
    \put(0.70516361,0.26345301){\makebox(0,0)[lt]{\lineheight{1.25}\smash{\begin{tabular}[t]{l}QFT\end{tabular}}}}%
    \put(0.86199682,0.26345301){\makebox(0,0)[lt]{\lineheight{1.25}\smash{\begin{tabular}[t]{l}VQE\end{tabular}}}}%
    \put(0.14001172,0.3031216){\makebox(0,0)[lt]{\lineheight{1.25}\smash{\begin{tabular}[t]{l}0\end{tabular}}}}%
    \put(0.12235334,0.35835635){\makebox(0,0)[lt]{\lineheight{1.25}\smash{\begin{tabular}[t]{l}20\end{tabular}}}}%
    \put(0.12235334,0.4135911){\makebox(0,0)[lt]{\lineheight{1.25}\smash{\begin{tabular}[t]{l}40\end{tabular}}}}%
    \put(0.12235334,0.46874215){\makebox(0,0)[lt]{\lineheight{1.25}\smash{\begin{tabular}[t]{l}60\end{tabular}}}}%
    \put(0.12235334,0.5239769){\makebox(0,0)[lt]{\lineheight{1.25}\smash{\begin{tabular}[t]{l}80\end{tabular}}}}%
    \put(0.10461126,0.57921165){\makebox(0,0)[lt]{\lineheight{1.25}\smash{\begin{tabular}[t]{l}100\end{tabular}}}}%
    \put(0.10461126,0.63436271){\makebox(0,0)[lt]{\lineheight{1.25}\smash{\begin{tabular}[t]{l}120\end{tabular}}}}%
    \put(0.10461126,0.68959746){\makebox(0,0)[lt]{\lineheight{1.25}\smash{\begin{tabular}[t]{l}140\end{tabular}}}}%
    \put(0.10461126,0.74474851){\makebox(0,0)[lt]{\lineheight{1.25}\smash{\begin{tabular}[t]{l}160\end{tabular}}}}%
    \put(0,0){\includegraphics[width=\unitlength,page=2]{pirot-counts.pdf}}%
    \put(0.1333166,0.12980166){\makebox(0,0)[lt]{\lineheight{1.25}\smash{\begin{tabular}[t]{l}Old Compiler: No Opt.\end{tabular}}}}%
    \put(0.58063436,0.12980166){\makebox(0,0)[lt]{\lineheight{1.25}\smash{\begin{tabular}[t]{l}Old Compiler: $RZ$ Opt.\end{tabular}}}}%
    \put(0.1333166,0.09004938){\makebox(0,0)[lt]{\lineheight{1.25}\smash{\begin{tabular}[t]{l}Ours: No Opt.\end{tabular}}}}%
    \put(0.58063436,0.09004938){\makebox(0,0)[lt]{\lineheight{1.25}\smash{\begin{tabular}[t]{l}Ours: Only Trailing Opt.\end{tabular}}}}%
    \put(0.1333166,0.0502971){\makebox(0,0)[lt]{\lineheight{1.25}\smash{\begin{tabular}[t]{l}Ours: Only from $RZ$ Opt.\end{tabular}}}}%
    \put(0.58063436,0.0502971){\makebox(0,0)[lt]{\lineheight{1.25}\smash{\begin{tabular}[t]{l}Ours: Only Up to $RZ$ Opt.\end{tabular}}}}%
    \put(0.1333166,0.01054482){\makebox(0,0)[lt]{\lineheight{1.25}\smash{\begin{tabular}[t]{l}Ours: Only $RX$ Opt.\end{tabular}}}}%
    \put(0.58063436,0.01054482){\makebox(0,0)[lt]{\lineheight{1.25}\smash{\begin{tabular}[t]{l}Ours: All Opt.\end{tabular}}}}%
    \put(0.49903758,0.20838564){\makebox(0,0)[lt]{\lineheight{1.25}\smash{\begin{tabular}[t]{l}Benchmark\end{tabular}}}}%
    \put(0.06720228,0.32362541){\rotatebox{90}{\makebox(0,0)[lt]{\lineheight{1.25}\smash{\begin{tabular}[t]{l}$R_\phi(\pi/2)$ Gate Count\end{tabular}}}}}%
  \end{picture}%
\endgroup%

%% file: fig/ms-hw-speedup-text.tex
%% Creator: Inkscape 1.1 (1:1.1+202106031931+af4d65493e), www.inkscape.org
%% Accompanies image file 'ms-hw-speedup.pdf' (pdf, eps, ps)
\begingroup%
  \makeatletter%
  \providecommand\color[2][]{%
    \errmessage{(Inkscape) Color is used for the text in Inkscape, but the package 'color.sty' is not loaded}%
    \renewcommand\color[2][]{}%
  }%
  \providecommand\transparent[1]{%
    \errmessage{(Inkscape) Transparency is used (non-zero) for the text in Inkscape, but the package 'transparent.sty' is not loaded}%
    \renewcommand\transparent[1]{}%
  }%
  \providecommand\rotatebox[2]{#2}%
  \newcommand*\fsize{\dimexpr\f@size pt\relax}%
  \newcommand*\lineheight[1]{\fontsize{\fsize}{#1\fsize}\selectfont}%
  \ifx\svgwidth\undefined%
    \setlength{\unitlength}{264.95433014bp}%
    \ifx\svgscale\undefined%
      \relax%
    \else%
      \setlength{\unitlength}{\unitlength * \real{\svgscale}}%
    \fi%
  \else%
    \setlength{\unitlength}{\svgwidth}%
  \fi%
  \global\let\svgwidth\undefined%
  \global\let\svgscale\undefined%
  \makeatother%
  \begin{picture}(1,0.82935555)%
    \lineheight{1}%
    \setlength\tabcolsep{0pt}%
    \put(0,0){\includegraphics[width=\unitlength,page=1]{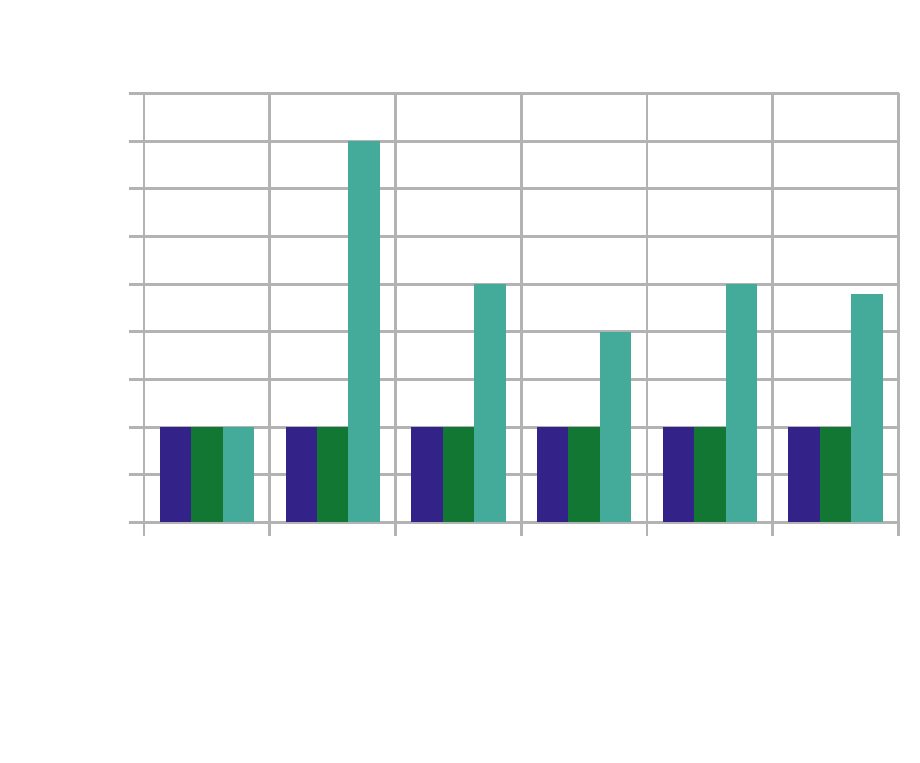}}%
    \put(0.1767412,0.19995721){\makebox(0,0)[lt]{\lineheight{1.25}\smash{\begin{tabular}[t]{l}GHZ\end{tabular}}}}%
    \put(0.32898256,0.19995721){\makebox(0,0)[lt]{\lineheight{1.25}\smash{\begin{tabular}[t]{l}BV\end{tabular}}}}%
    \put(0.43457794,0.19995721){\makebox(0,0)[lt]{\lineheight{1.25}\smash{\begin{tabular}[t]{l}Grover\end{tabular}}}}%
    \put(0.59109875,0.19995721){\makebox(0,0)[lt]{\lineheight{1.25}\smash{\begin{tabular}[t]{l}QFT\end{tabular}}}}%
    \put(0.72354766,0.19995721){\makebox(0,0)[lt]{\lineheight{1.25}\smash{\begin{tabular}[t]{l}VQE\end{tabular}}}}%
    \put(0.83192468,0.19995721){\makebox(0,0)[lt]{\lineheight{1.25}\smash{\begin{tabular}[t]{l}Average\end{tabular}}}}%
    \put(0.10720017,0.24703113){\makebox(0,0)[lt]{\lineheight{1.25}\smash{\begin{tabular}[t]{l}0\end{tabular}}}}%
    \put(0.07328555,0.29881245){\makebox(0,0)[lt]{\lineheight{1.25}\smash{\begin{tabular}[t]{l}0.5\end{tabular}}}}%
    \put(0.10720017,0.35059377){\makebox(0,0)[lt]{\lineheight{1.25}\smash{\begin{tabular}[t]{l}1\end{tabular}}}}%
    \put(0.07328555,0.40237509){\makebox(0,0)[lt]{\lineheight{1.25}\smash{\begin{tabular}[t]{l}1.5\end{tabular}}}}%
    \put(0.10720017,0.4542634){\makebox(0,0)[lt]{\lineheight{1.25}\smash{\begin{tabular}[t]{l}2\end{tabular}}}}%
    \put(0.07328555,0.50604472){\makebox(0,0)[lt]{\lineheight{1.25}\smash{\begin{tabular}[t]{l}2.5\end{tabular}}}}%
    \put(0.10720017,0.55782604){\makebox(0,0)[lt]{\lineheight{1.25}\smash{\begin{tabular}[t]{l}3\end{tabular}}}}%
    \put(0.07328555,0.60960736){\makebox(0,0)[lt]{\lineheight{1.25}\smash{\begin{tabular}[t]{l}3.5\end{tabular}}}}%
    \put(0.10720017,0.66138868){\makebox(0,0)[lt]{\lineheight{1.25}\smash{\begin{tabular}[t]{l}4\end{tabular}}}}%
    \put(0.07328555,0.71317){\makebox(0,0)[lt]{\lineheight{1.25}\smash{\begin{tabular}[t]{l}4.5\end{tabular}}}}%
    \put(0.20370172,0.37509361){\rotatebox{90}{\makebox(0,0)[lt]{\lineheight{1.25}\smash{\begin{tabular}[t]{l}\footnotesize 1.00\end{tabular}}}}}%
    \put(0.34043008,0.37509361){\rotatebox{90}{\makebox(0,0)[lt]{\lineheight{1.25}\smash{\begin{tabular}[t]{l}\footnotesize 1.00\end{tabular}}}}}%
    \put(0.47705146,0.37509361){\rotatebox{90}{\makebox(0,0)[lt]{\lineheight{1.25}\smash{\begin{tabular}[t]{l}\footnotesize 1.00\end{tabular}}}}}%
    \put(0.61367284,0.37509361){\rotatebox{90}{\makebox(0,0)[lt]{\lineheight{1.25}\smash{\begin{tabular}[t]{l}\footnotesize 1.00\end{tabular}}}}}%
    \put(0.7504012,0.37509361){\rotatebox{90}{\makebox(0,0)[lt]{\lineheight{1.25}\smash{\begin{tabular}[t]{l}\footnotesize 1.00\end{tabular}}}}}%
    \put(0.88702257,0.37509361){\rotatebox{90}{\makebox(0,0)[lt]{\lineheight{1.25}\smash{\begin{tabular}[t]{l}\footnotesize 1.00\end{tabular}}}}}%
    \put(0.23783032,0.37509361){\rotatebox{90}{\makebox(0,0)[lt]{\lineheight{1.25}\smash{\begin{tabular}[t]{l}\footnotesize 1.00\end{tabular}}}}}%
    \put(0.37455868,0.37509361){\rotatebox{90}{\makebox(0,0)[lt]{\lineheight{1.25}\smash{\begin{tabular}[t]{l}\footnotesize 1.00\end{tabular}}}}}%
    \put(0.51118006,0.37509361){\rotatebox{90}{\makebox(0,0)[lt]{\lineheight{1.25}\smash{\begin{tabular}[t]{l}\footnotesize 1.00\end{tabular}}}}}%
    \put(0.64780143,0.37509361){\rotatebox{90}{\makebox(0,0)[lt]{\lineheight{1.25}\smash{\begin{tabular}[t]{l}\footnotesize 1.00\end{tabular}}}}}%
    \put(0.7845298,0.37509361){\rotatebox{90}{\makebox(0,0)[lt]{\lineheight{1.25}\smash{\begin{tabular}[t]{l}\footnotesize 1.00\end{tabular}}}}}%
    \put(0.92125816,0.37509361){\rotatebox{90}{\makebox(0,0)[lt]{\lineheight{1.25}\smash{\begin{tabular}[t]{l}\footnotesize 1.00\end{tabular}}}}}%
    \put(0.2720659,0.37509361){\rotatebox{90}{\makebox(0,0)[lt]{\lineheight{1.25}\smash{\begin{tabular}[t]{l}\footnotesize 1.00\end{tabular}}}}}%
    \put(0.40868728,0.68588852){\rotatebox{90}{\makebox(0,0)[lt]{\lineheight{1.25}\smash{\begin{tabular}[t]{l}\footnotesize 4.00\end{tabular}}}}}%
    \put(0.54530866,0.53054456){\rotatebox{90}{\makebox(0,0)[lt]{\lineheight{1.25}\smash{\begin{tabular}[t]{l}\footnotesize 2.50\end{tabular}}}}}%
    \put(0.68203702,0.47876324){\rotatebox{90}{\makebox(0,0)[lt]{\lineheight{1.25}\smash{\begin{tabular}[t]{l}\footnotesize 2.00\end{tabular}}}}}%
    \put(0.81865839,0.53054456){\rotatebox{90}{\makebox(0,0)[lt]{\lineheight{1.25}\smash{\begin{tabular}[t]{l}\footnotesize 2.50\end{tabular}}}}}%
    \put(0.95538676,0.5201669){\rotatebox{90}{\makebox(0,0)[lt]{\lineheight{1.25}\smash{\begin{tabular}[t]{l}\footnotesize 2.40\end{tabular}}}}}%
    \put(0,0){\includegraphics[width=\unitlength,page=2]{ms-hw-speedup.pdf}}%
    \put(0.09735744,0.06836418){\makebox(0,0)[lt]{\lineheight{1.25}\smash{\begin{tabular}[t]{l}Old Compiler\end{tabular}}}}%
    \put(0.69102386,0.06836418){\makebox(0,0)[lt]{\lineheight{1.25}\smash{\begin{tabular}[t]{l}Our Compiler\end{tabular}}}}%
    \put(0.09735744,0.02096929){\makebox(0,0)[lt]{\lineheight{1.25}\smash{\begin{tabular}[t]{l}HW Upgrade: Fully Conn.\end{tabular}}}}%
    \put(0.46218038,0.14207767){\makebox(0,0)[lt]{\lineheight{1.25}\smash{\begin{tabular}[t]{l}Benchmark\end{tabular}}}}%
    \put(0.05317214,0.27463357){\rotatebox{90}{\makebox(0,0)[lt]{\lineheight{1.25}\smash{\begin{tabular}[t]{l}$XX(\pi/4)$ Count Reduction\end{tabular}}}}}%
  \end{picture}%
\endgroup%

%% file: fig/pirot-hw-speedup-text.tex
%% Creator: Inkscape 1.1 (1:1.1+202106031931+af4d65493e), www.inkscape.org
%% Accompanies image file 'pirot-hw-speedup.pdf' (pdf, eps, ps)
\begingroup%
  \makeatletter%
  \providecommand\color[2][]{%
    \errmessage{(Inkscape) Color is used for the text in Inkscape, but the package 'color.sty' is not loaded}%
    \renewcommand\color[2][]{}%
  }%
  \providecommand\transparent[1]{%
    \errmessage{(Inkscape) Transparency is used (non-zero) for the text in Inkscape, but the package 'transparent.sty' is not loaded}%
    \renewcommand\transparent[1]{}%
  }%
  \providecommand\rotatebox[2]{#2}%
  \newcommand*\fsize{\dimexpr\f@size pt\relax}%
  \newcommand*\lineheight[1]{\fontsize{\fsize}{#1\fsize}\selectfont}%
  \ifx\svgwidth\undefined%
    \setlength{\unitlength}{468.36850598bp}%
    \ifx\svgscale\undefined%
      \relax%
    \else%
      \setlength{\unitlength}{\unitlength * \real{\svgscale}}%
    \fi%
  \else%
    \setlength{\unitlength}{\svgwidth}%
  \fi%
  \global\let\svgwidth\undefined%
  \global\let\svgscale\undefined%
  \makeatother%
  \begin{picture}(1,0.76532629)%
    \lineheight{1}%
    \setlength\tabcolsep{0pt}%
    \put(0,0){\includegraphics[width=\unitlength,page=1]{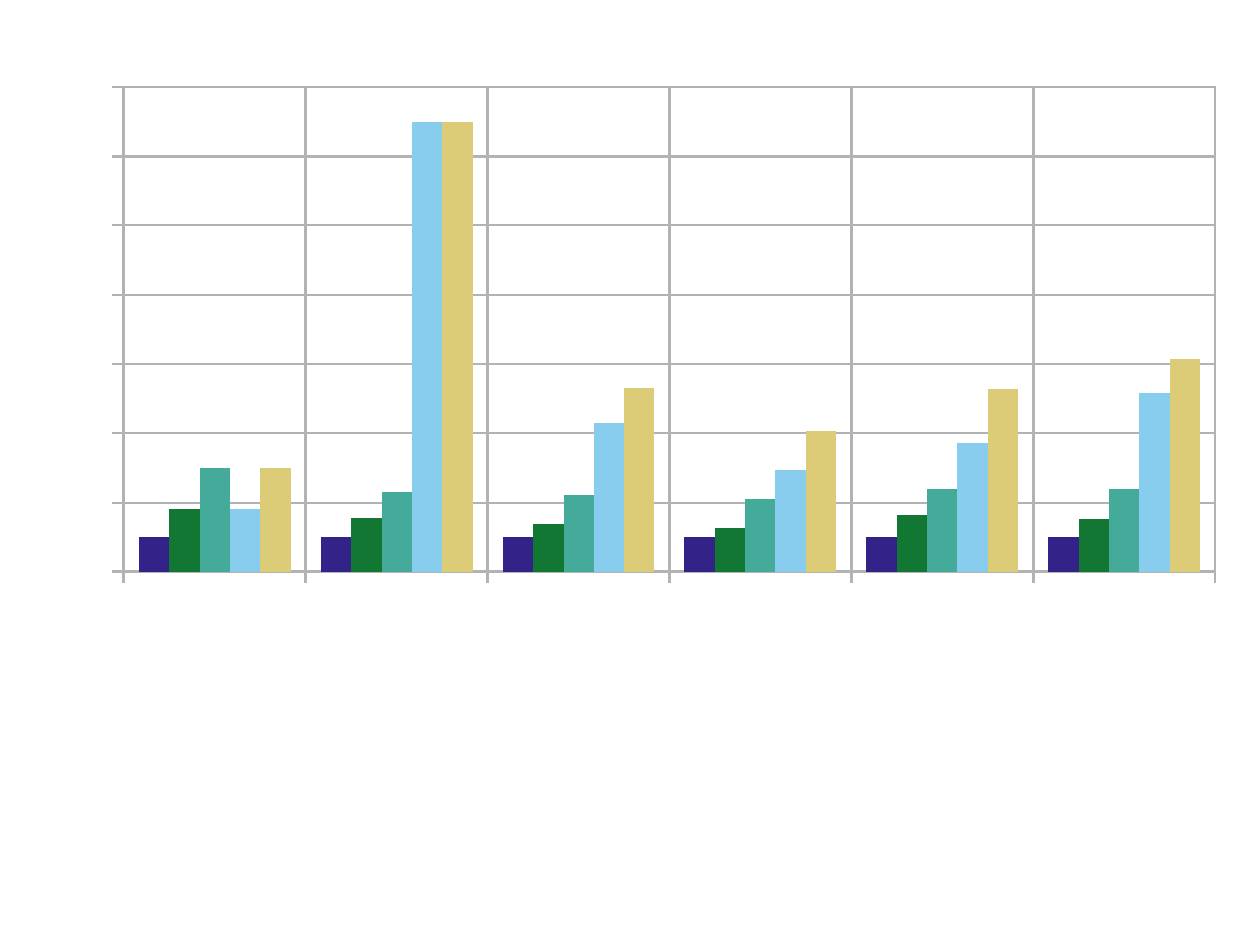}}%
    \put(0.14537312,0.2697452){\makebox(0,0)[lt]{\lineheight{1.25}\smash{\begin{tabular}[t]{l}GHZ\end{tabular}}}}%
    \put(0.30049023,0.2697452){\makebox(0,0)[lt]{\lineheight{1.25}\smash{\begin{tabular}[t]{l}BV\end{tabular}}}}%
    \put(0.42921988,0.2697452){\makebox(0,0)[lt]{\lineheight{1.25}\smash{\begin{tabular}[t]{l}Grover\end{tabular}}}}%
    \put(0.58681837,0.2697452){\makebox(0,0)[lt]{\lineheight{1.25}\smash{\begin{tabular}[t]{l}QFT\end{tabular}}}}%
    \put(0.73073897,0.2697452){\makebox(0,0)[lt]{\lineheight{1.25}\smash{\begin{tabular}[t]{l}VQE\end{tabular}}}}%
    \put(0.86104218,0.2697452){\makebox(0,0)[lt]{\lineheight{1.25}\smash{\begin{tabular}[t]{l}Average\end{tabular}}}}%
    \put(0.07153665,0.29716153){\makebox(0,0)[lt]{\lineheight{1.25}\smash{\begin{tabular}[t]{l}0\end{tabular}}}}%
    \put(0.07153665,0.35284149){\makebox(0,0)[lt]{\lineheight{1.25}\smash{\begin{tabular}[t]{l}2\end{tabular}}}}%
    \put(0.07153665,0.40858198){\makebox(0,0)[lt]{\lineheight{1.25}\smash{\begin{tabular}[t]{l}4\end{tabular}}}}%
    \put(0.07153665,0.46426194){\makebox(0,0)[lt]{\lineheight{1.25}\smash{\begin{tabular}[t]{l}6\end{tabular}}}}%
    \put(0.07153665,0.52000242){\makebox(0,0)[lt]{\lineheight{1.25}\smash{\begin{tabular}[t]{l}8\end{tabular}}}}%
    \put(0.05876657,0.57562186){\makebox(0,0)[lt]{\lineheight{1.25}\smash{\begin{tabular}[t]{l}10\end{tabular}}}}%
    \put(0.05876657,0.63136234){\makebox(0,0)[lt]{\lineheight{1.25}\smash{\begin{tabular}[t]{l}12\end{tabular}}}}%
    \put(0.05876657,0.68710283){\makebox(0,0)[lt]{\lineheight{1.25}\smash{\begin{tabular}[t]{l}14\end{tabular}}}}%
    \put(0.13115052,0.33928463){\rotatebox{90}{\makebox(0,0)[lt]{\lineheight{1.25}\smash{\begin{tabular}[t]{l}\footnotesize 1.00\end{tabular}}}}}%
    \put(0.27749198,0.33928463){\rotatebox{90}{\makebox(0,0)[lt]{\lineheight{1.25}\smash{\begin{tabular}[t]{l}\footnotesize 1.00\end{tabular}}}}}%
    \put(0.42383344,0.33928463){\rotatebox{90}{\makebox(0,0)[lt]{\lineheight{1.25}\smash{\begin{tabular}[t]{l}\footnotesize 1.00\end{tabular}}}}}%
    \put(0.57011439,0.33928463){\rotatebox{90}{\makebox(0,0)[lt]{\lineheight{1.25}\smash{\begin{tabular}[t]{l}\footnotesize 1.00\end{tabular}}}}}%
    \put(0.71645585,0.33928463){\rotatebox{90}{\makebox(0,0)[lt]{\lineheight{1.25}\smash{\begin{tabular}[t]{l}\footnotesize 1.00\end{tabular}}}}}%
    \put(0.86279731,0.33928463){\rotatebox{90}{\makebox(0,0)[lt]{\lineheight{1.25}\smash{\begin{tabular}[t]{l}\footnotesize 1.00\end{tabular}}}}}%
    \put(0.15554076,0.36155662){\rotatebox{90}{\makebox(0,0)[lt]{\lineheight{1.25}\smash{\begin{tabular}[t]{l}\footnotesize 1.80\end{tabular}}}}}%
    \put(0.30188222,0.35489923){\rotatebox{90}{\makebox(0,0)[lt]{\lineheight{1.25}\smash{\begin{tabular}[t]{l}\footnotesize 1.56\end{tabular}}}}}%
    \put(0.44822369,0.3500575){\rotatebox{90}{\makebox(0,0)[lt]{\lineheight{1.25}\smash{\begin{tabular}[t]{l}\footnotesize 1.39\end{tabular}}}}}%
    \put(0.59450463,0.34624463){\rotatebox{90}{\makebox(0,0)[lt]{\lineheight{1.25}\smash{\begin{tabular}[t]{l}\footnotesize 1.25\end{tabular}}}}}%
    \put(0.74084609,0.3564728){\rotatebox{90}{\makebox(0,0)[lt]{\lineheight{1.25}\smash{\begin{tabular}[t]{l}\footnotesize 1.62\end{tabular}}}}}%
    \put(0.88718756,0.35380984){\rotatebox{90}{\makebox(0,0)[lt]{\lineheight{1.25}\smash{\begin{tabular}[t]{l}\footnotesize 1.52\end{tabular}}}}}%
    \put(0.17993101,0.39496459){\rotatebox{90}{\makebox(0,0)[lt]{\lineheight{1.25}\smash{\begin{tabular}[t]{l}\footnotesize 3.00\end{tabular}}}}}%
    \put(0.32627247,0.37529504){\rotatebox{90}{\makebox(0,0)[lt]{\lineheight{1.25}\smash{\begin{tabular}[t]{l}\footnotesize 2.29\end{tabular}}}}}%
    \put(0.47261393,0.37317678){\rotatebox{90}{\makebox(0,0)[lt]{\lineheight{1.25}\smash{\begin{tabular}[t]{l}\footnotesize 2.22\end{tabular}}}}}%
    \put(0.61889487,0.37057435){\rotatebox{90}{\makebox(0,0)[lt]{\lineheight{1.25}\smash{\begin{tabular}[t]{l}\footnotesize 2.13\end{tabular}}}}}%
    \put(0.76523634,0.3774133){\rotatebox{90}{\makebox(0,0)[lt]{\lineheight{1.25}\smash{\begin{tabular}[t]{l}\footnotesize 2.37\end{tabular}}}}}%
    \put(0.9115778,0.37826061){\rotatebox{90}{\makebox(0,0)[lt]{\lineheight{1.25}\smash{\begin{tabular}[t]{l}\footnotesize 2.40\end{tabular}}}}}%
    \put(0.20432125,0.36155662){\rotatebox{90}{\makebox(0,0)[lt]{\lineheight{1.25}\smash{\begin{tabular}[t]{l}\footnotesize 1.80\end{tabular}}}}}%
    \put(0.35066271,0.67348544){\rotatebox{90}{\makebox(0,0)[lt]{\lineheight{1.25}\smash{\begin{tabular}[t]{l}\footnotesize 13.00\end{tabular}}}}}%
    \put(0.49700418,0.43091448){\rotatebox{90}{\makebox(0,0)[lt]{\lineheight{1.25}\smash{\begin{tabular}[t]{l}\footnotesize 4.29\end{tabular}}}}}%
    \put(0.64328512,0.3930279){\rotatebox{90}{\makebox(0,0)[lt]{\lineheight{1.25}\smash{\begin{tabular}[t]{l}\footnotesize 2.93\end{tabular}}}}}%
    \put(0.78962658,0.41536041){\rotatebox{90}{\makebox(0,0)[lt]{\lineheight{1.25}\smash{\begin{tabular}[t]{l}\footnotesize 3.73\end{tabular}}}}}%
    \put(0.93596804,0.45488107){\rotatebox{90}{\makebox(0,0)[lt]{\lineheight{1.25}\smash{\begin{tabular}[t]{l}\footnotesize 5.15\end{tabular}}}}}%
    \put(0.22871149,0.39496459){\rotatebox{90}{\makebox(0,0)[lt]{\lineheight{1.25}\smash{\begin{tabular}[t]{l}\footnotesize 3.00\end{tabular}}}}}%
    \put(0.37505296,0.67348544){\rotatebox{90}{\makebox(0,0)[lt]{\lineheight{1.25}\smash{\begin{tabular}[t]{l}\footnotesize 13.00\end{tabular}}}}}%
    \put(0.5213339,0.45954125){\rotatebox{90}{\makebox(0,0)[lt]{\lineheight{1.25}\smash{\begin{tabular}[t]{l}\footnotesize 5.32\end{tabular}}}}}%
    \put(0.66767536,0.42419657){\rotatebox{90}{\makebox(0,0)[lt]{\lineheight{1.25}\smash{\begin{tabular}[t]{l}\footnotesize 4.05\end{tabular}}}}}%
    \put(0.81401683,0.4580282){\rotatebox{90}{\makebox(0,0)[lt]{\lineheight{1.25}\smash{\begin{tabular}[t]{l}\footnotesize 5.26\end{tabular}}}}}%
    \put(0.96035829,0.48205532){\rotatebox{90}{\makebox(0,0)[lt]{\lineheight{1.25}\smash{\begin{tabular}[t]{l}\footnotesize 6.13\end{tabular}}}}}%
    \put(0,0){\includegraphics[width=\unitlength,page=2]{pirot-hw-speedup.pdf}}%
    \put(0.39738546,0.18344126){\makebox(0,0)[lt]{\lineheight{1.25}\smash{\begin{tabular}[t]{l}Old Compiler\end{tabular}}}}%
    \put(0.39738546,0.14313381){\makebox(0,0)[lt]{\lineheight{1.25}\smash{\begin{tabular}[t]{l}Our Compiler\end{tabular}}}}%
    \put(0.39738546,0.10282636){\makebox(0,0)[lt]{\lineheight{1.25}\smash{\begin{tabular}[t]{l}HW Upgrade: Parallel 1Q Ops\end{tabular}}}}%
    \put(0.39738546,0.06251891){\makebox(0,0)[lt]{\lineheight{1.25}\smash{\begin{tabular}[t]{l}HW Upgrade: Fully Conn.\end{tabular}}}}%
    \put(0.39738546,0.02221146){\makebox(0,0)[lt]{\lineheight{1.25}\smash{\begin{tabular}[t]{l}HW Upgrade: Both\end{tabular}}}}%
    \put(0.49264661,0.22719845){\makebox(0,0)[lt]{\lineheight{1.25}\smash{\begin{tabular}[t]{l}Benchmark\end{tabular}}}}%
    \put(0.03007928,0.32863281){\rotatebox{90}{\makebox(0,0)[lt]{\lineheight{1.25}\smash{\begin{tabular}[t]{l}$R_\phi(\pi/2)$ Cycle Reduction\end{tabular}}}}}%
  \end{picture}%
\endgroup%

%% file: sections/05-adaption-other-hw.tex
\section{Adaption to Other Hardware}
%While our compiler was designed for a particular ion trap machine, our optimization strategies work on other hardware with a similar gateset. For example, the QSCOUT ion trap quantum testbed operated by Sandia National Laboratories natively supports $R_\phi(\theta)$ and $XX(\alpha)$ (Equations \ref{eq:rot-exp} and \ref{eq:ms-exp} in Section \ref{sec:iontrap}). IonQ hardware also supports .
While our compiler was designed for a particular ion trap machine, our backend can be adapted for any hardware with a similar gateset. For example, some IonQ hardware and the QSCOUT ion trap quantum testbed operated by Sandia National Laboratories natively support $R_\phi(\theta)$ and $XX(\alpha)$ gates (Equations \ref{eq:rot-exp} and \ref{eq:xx-gate-mat} in Section \ref{sec:iontrap}) \cite{debnath_demonstration_2016,morrison_just_2020}. On these machines, the angle $\theta$ in $R_\phi(\theta)$ is not limited to $\pi/2$ as in the GTRI testbed, so our numerical optimizer approach explained in Section \ref{sec:1q-pass} is not necessary, as there is a closed form solution for decomposing arbitrary single-qubit unitaries into two $R_\phi(\theta)$ gates \cite{maslov_basic_2017}. However, our approach of removing unnecessary $RZ$ gates and commuting $RX$ gates described in the remainder of Section \ref{sec:1q-pass} would likely still reduce gate count.

%\aja{tom's question: Do other systems use $\pi / 2$ for calibration?}

Still, it is a common choice to restrict $\theta$ in $R_\phi(\theta)$ to limit the set of calibrations. For example, a recent ion trap quantum computer developed by Honeywell supports $R_\phi(\theta)$ with $\theta$ constrained to $\pi$ or $\pi/2$, much like the $\theta=\pi/2$ restriction on the GTRI testbed \cite{pino_demonstration_2021}. The numerical optimization strategy described in Section \ref{sec:1q-pass} could be adapted by running the up-to-Z optimizer on all seven combinations of zero through two $R_\phi(\theta)$ rotations with each individual angle $\theta \in \{\pi,\pi/2\}$, all followed by a Z-rotation.
%\aja{does this make sense? the paper says they decompose any unitary into two $R$ gates and then a Z-rotation.}
The machine has the entangling gate $ZZ(\alpha) = \exp(-i\alpha(Z \otimes Z))$, analogous to the $XX(\alpha)$ gate. $ZZ(\alpha)$ commutes with $RZ(\theta)$, so our approach for commuting $RX(\theta)$ around $XX(\alpha)$ gates could be adapted to commute $RZ(\theta)$ gates around $ZZ(\alpha)$ gates instead. Despite these similarities, a compiler for the Honeywell machine would need to understand the multiple transport operations supported; our backend does not consider transport operations, since the testbed control software infers them automatically.

The current domain-specific configuration of the GTRI testbed could also be supported, as well as other hardware with global operations in general, but this would require adding new instructions to XACC IR targeting all qubits. Moreover, our IR transformations described in Sections \ref{sec:2q-pass} and \ref{sec:1q-pass} will not be useful on hardware with only global operations, since one would not typically program them using a typical quantum circuit. However, backends for any hardware whose native operations support typical quantum circuits benefit from the existing high-level circuit optimizations already present in QCOR \cite{nguyen_extending_2020}.
%\aja{jeff's point: end this on a more positive note by saying that IR optimizations in QCOR apply to any hardware}

%\revone{The general idea of the paper is good, but it relies heavily on the backend staying up to date with the hardware system and it is unclear how generally applied such a backend will be. There is evidence for this in the fact that the target testbed hardware was modified already in a way that it cannot execute the benchmark circuits for the paper. Some additional commentary about whether or not there is any generality in the XACC backend would be helpful. It may be that it is not general, or that parts of the backend will only need minor changes to be adapted to a different system. It would be helpful to the reader to hear the authors vision for this given that the current state of quantum hardware is changing rapidly.}\aja{Added this section to try to address this}

%% file: sections/06-rel-work.tex
\section{Related Work}

Martinez et al. \cite{martinez_compiling_2016} discuss techniques for compiling arbitrary multi-qubit gates to ion-trap architectures supporting collective rotations of the whole qubit register about any axis ($C(\theta, \phi)$ gate), single qubit rotations around the $Z$ axis ($Z_n(\theta)$ gate), and MS gates. They show how any multi-qubit gate can be decomposed into a sequence of single-qubit local unitaries and MS gates. Our work employs this decomposition concretely in the XACC backend restricted to the single-qubit and two-qubit case. Maslov \cite{maslov_basic_2017} proposes a generic architecture for optimizing compilers targeting ion-trap quantum computers. QCOR does not match the architecture exactly, but instead provides a framework for describing and applying relevant optimizations. Our implementation does not support the peep-hole optimizations proposed in the paper.

QFAST \cite{younis_qfast_2021} innovates in circuit synthesis by presenting a novel representation of circuits allowing them to use numerical optimization to replace expensive searches over circuit structures. They use a bottom-up approach which stems from a special encoding that enables them to find better building blocks for circuit synthesis. QGo \cite{wu_qgo_2021} is a quantum circuit optimization framework that aims to be scalable (optimizing circuits containing 60+ qubits in a reasonable amount of time). It is able to achieve this performance by employing a partitioning scheme in which the circuit is broken down into smaller components, independently optimized, and then composed together.

Lu et al. \cite{lu_global_2019} discuss a scalable scheme for implementing a global multi-qubit entangling gate which can potentially lead to exponential speedups as compared to a circuit decomposition involving single- and two-qubit entangling gates. The GTRI testbed is currently configured for specific multi-qubit global entangling operations \cite{rajakumar_generating_2020}, but we have not considered them in this work.

Gokhale et al. \cite{gheorghiu_quantum_2018} have implemented a compiler that exploits pulse-level optimizations without resorting to Quantum Optimal Control (QOC) approaches, thereby bypassing the experimental barrier of measuring and maintaining the Hamiltonian of a quantum system. They are able to achieve about $1.6{\times}$ error reductions and $2{\times}$ speedups on near-time quantum algorithms run using the OpenPulse interface on IBM's quantum computers. Our work does not currently employ these optimizations, which we leave as future work.

Pino et al. \cite{pino_demonstration_2021} demonstrate a quantum computer also based on an ion trap, but contrary to our assumptions in this work about future GTRI testbed upgrades (i.e., per-ion beams), their design transports ions through shared beams to perform gates. Ion swapping operations help provide full qubit connectivity, and multiple beams provide support for parallel operations. The authors briefly mention a compiler that performs qubit mapping such that it minimizes the number of native transport operations, which may include linear ion movement, swapping ion order, and splitting or combining ion crystals, but they do not go into detail on the compiler design. Since our assumptions about future hardware upgrades to the GTRI testbed are only guesses, future work should consider minimizing the number of these transport operations in addition to native gate count, our primary consideration in this work.

The TILT (Trapped-Ion Linear Tape) architecture for ion trap hardware proposed by Wu et al. consists of a stationary set of lasers targeting a subset of the ions in a single ion chain~\cite{wu_tilt_2021}. Compared to the previously proposed QCCD (Quantum Charge-Coupled Device) architecture~\cite{kielpinski_architecture_2002}, TILT offers simpler hardware and avoids expensive shuttling operations. The authors detail LinQ, their compiler framework designed for TILT hardware, which employs two heuristic-based algorithms: one for inserting SWAP gates, and another for transport operations. The first algorithm facilitates two-qubit gates between qubits not within the ``execution zone'' of the lasers, and the second attempts to avoid unnecessary tape movement, which may introduce qubit noise. Their discussion of LinQ further emphasizes the importance of future work on our backend minimizing the number of transport operations. Additionally, it is not mentioned if their compiler employs the optimizations employed in Section \ref{sec:1q-pass}. Future work should investigate how to apply these optimizations to TILT systems.

%% file: sections/07-conclusions.tex
\section{Conclusion}
This work details efforts to add a new ion trap backend to the XACC quantum toolchain as well as a demonstration of multi-level optimization strategies to provide algorithmic optimizations at the language, IR, and backend levels. As a demonstration of this strategy, our implementation allows heterogeneous quantum--C++ programs to be compiled and optimized to use fewer physical operations, along with a basic simulation functionality using the testbed's existing development tools. 

Future work in this space would look to extend this programming environment to support further optimizations as well as testing with the quantum hardware instead of the simulated environment. We would first consider optimizations such as parallel two-qubit gates, non-$R_\phi(\pi/2)$ operations, and multi-qubit optimizations to improve the fidelity of generated quantum circuits.

%% file: paper.bbl
\begin{thebibliography}{10}
\providecommand{\url}[1]{#1}
\csname url@samestyle\endcsname
\providecommand{\newblock}{\relax}
\providecommand{\bibinfo}[2]{#2}
\providecommand{\BIBentrySTDinterwordspacing}{\spaceskip=0pt\relax}
\providecommand{\BIBentryALTinterwordstretchfactor}{4}
\providecommand{\BIBentryALTinterwordspacing}{\spaceskip=\fontdimen2\font plus
\BIBentryALTinterwordstretchfactor\fontdimen3\font minus
  \fontdimen4\font\relax}
\providecommand{\BIBforeignlanguage}[2]{{%
\expandafter\ifx\csname l@#1\endcsname\relax
\typeout{** WARNING: IEEEtran.bst: No hyphenation pattern has been}%
\typeout{** loaded for the language `#1'. Using the pattern for}%
\typeout{** the default language instead.}%
\else
\language=\csname l@#1\endcsname
\fi
#2}}
\providecommand{\BIBdecl}{\relax}
\BIBdecl

\bibitem{nielsen_quantum_2010}
M.~A. Nielsen and I.~L. Chuang, \emph{\BIBforeignlanguage{English}{Quantum
  {Computation} and {Quantum} {Information}: 10th {Anniversary} {Edition}}},
  1st~ed.\hskip 1em plus 0.5em minus 0.4em\relax Cambridge ; New York:
  Cambridge University Press, Dec. 2010.

\bibitem{preskill_quantum_2018}
\BIBentryALTinterwordspacing
J.~Preskill, ``\BIBforeignlanguage{en-GB}{Quantum {Computing} in the {NISQ} era
  and beyond},'' \emph{\BIBforeignlanguage{en-GB}{Quantum}}, vol.~2, p.~79,
  Aug. 2018, publisher: Verein zur Förderung des Open Access Publizierens in
  den Quantenwissenschaften. [Online]. Available:
  \url{https://quantum-journal.org/papers/q-2018-08-06-79/}
\BIBentrySTDinterwordspacing

\bibitem{chong_programming_2017}
\BIBentryALTinterwordspacing
F.~T. Chong, D.~Franklin, and M.~Martonosi,
  ``\BIBforeignlanguage{en}{Programming languages and compiler design for
  realistic quantum hardware},'' \emph{\BIBforeignlanguage{en}{Nature}}, vol.
  549, no. 7671, pp. 180--187, Sep. 2017, number: 7671 Publisher: Nature
  Publishing Group. [Online]. Available:
  \url{https://www.nature.com/articles/nature23459}
\BIBentrySTDinterwordspacing

\bibitem{mccaskey_xacc_2020}
\BIBentryALTinterwordspacing
A.~J. McCaskey, D.~I. Lyakh, E.~F. Dumitrescu, S.~S. Powers, and T.~S. Humble,
  ``\BIBforeignlanguage{en}{{XACC}: a system-level software infrastructure for
  heterogeneous quantum–classical computing},''
  \emph{\BIBforeignlanguage{en}{Quantum Science and Technology}}, vol.~5,
  no.~2, p. 024002, Feb. 2020, publisher: IOP Publishing. [Online]. Available:
  \url{https://doi.org/10.1088/2058-9565/ab6bf6}
\BIBentrySTDinterwordspacing

\bibitem{herold_universal_2016}
\BIBentryALTinterwordspacing
C.~D. Herold, S.~D. Fallek, J.~T. Merrill, A.~M. Meier, K.~R. Brown, C.~E.
  Volin, and J.~M. Amini, ``\BIBforeignlanguage{en}{Universal control of ion
  qubits in a scalable microfabricated planar trap},''
  \emph{\BIBforeignlanguage{en}{New Journal of Physics}}, vol.~18, no.~2, p.
  023048, Feb. 2016, publisher: IOP Publishing. [Online]. Available:
  \url{https://doi.org/10.1088/1367-2630/18/2/023048}
\BIBentrySTDinterwordspacing

\bibitem{mintz_qcor_2020}
\BIBentryALTinterwordspacing
T.~M. Mintz, A.~J. McCaskey, E.~F. Dumitrescu, S.~V. Moore, S.~Powers, and
  P.~Lougovski, ``{QCOR}: {A} {Language} {Extension} {Specification} for the
  {Heterogeneous} {Quantum}-{Classical} {Model} of {Computation},'' \emph{ACM
  Journal on Emerging Technologies in Computing Systems}, vol.~16, no.~2, pp.
  22:1--22:17, Mar. 2020. [Online]. Available:
  \url{https://doi.org/10.1145/3380964}
\BIBentrySTDinterwordspacing

\bibitem{maslov_basic_2017}
\BIBentryALTinterwordspacing
D.~Maslov, ``\BIBforeignlanguage{en}{Basic circuit compilation techniques for
  an ion-trap quantum machine},'' \emph{\BIBforeignlanguage{en}{New Journal of
  Physics}}, vol.~19, no.~2, p. 023035, Feb. 2017, publisher: IOP Publishing.
  [Online]. Available: \url{https://doi.org/10.1088/1367-2630/aa5e47}
\BIBentrySTDinterwordspacing

\bibitem{molmer_multiparticle_1999}
\BIBentryALTinterwordspacing
K.~Mølmer and A.~Sørensen, ``Multiparticle {Entanglement} of {Hot} {Trapped}
  {Ions},'' \emph{Physical Review Letters}, vol.~82, no.~9, pp. 1835--1838,
  Mar. 1999, publisher: American Physical Society. [Online]. Available:
  \url{https://link.aps.org/doi/10.1103/PhysRevLett.82.1835}
\BIBentrySTDinterwordspacing

\bibitem{sorensen_quantum_1999}
\BIBentryALTinterwordspacing
A.~Sørensen and K.~Mølmer, ``Quantum {Computation} with {Ions} in {Thermal}
  {Motion},'' \emph{Physical Review Letters}, vol.~82, no.~9, pp. 1971--1974,
  Mar. 1999, publisher: American Physical Society. [Online]. Available:
  \url{https://link.aps.org/doi/10.1103/PhysRevLett.82.1971}
\BIBentrySTDinterwordspacing

\bibitem{debnath_demonstration_2016}
\BIBentryALTinterwordspacing
S.~Debnath, N.~M. Linke, C.~Figgatt, K.~A. Landsman, K.~Wright, and C.~Monroe,
  ``\BIBforeignlanguage{en}{Demonstration of a small programmable quantum
  computer with atomic qubits},'' \emph{\BIBforeignlanguage{en}{Nature}}, vol.
  536, no. 7614, pp. 63--66, Aug. 2016, number: 7614 Publisher: Nature
  Publishing Group. [Online]. Available:
  \url{https://www.nature.com/articles/nature18648}
\BIBentrySTDinterwordspacing

\bibitem{brylinski_universal_2001}
\BIBentryALTinterwordspacing
J.-L. Brylinski and R.~Brylinski, ``Universal quantum gates,''
  \emph{arXiv:quant-ph/0108062}, Aug. 2001, arXiv: quant-ph/0108062. [Online].
  Available: \url{http://arxiv.org/abs/quant-ph/0108062}
\BIBentrySTDinterwordspacing

\bibitem{bremner_practical_2002}
\BIBentryALTinterwordspacing
M.~J. Bremner, C.~M. Dawson, J.~L. Dodd, A.~Gilchrist, A.~W. Harrow,
  D.~Mortimer, M.~A. Nielsen, and T.~J. Osborne, ``Practical {Scheme} for
  {Quantum} {Computation} with {Any} {Two}-{Qubit} {Entangling} {Gate},''
  \emph{Physical Review Letters}, vol.~89, no.~24, p. 247902, Nov. 2002,
  publisher: American Physical Society. [Online]. Available:
  \url{https://link.aps.org/doi/10.1103/PhysRevLett.89.247902}
\BIBentrySTDinterwordspacing

\bibitem{figgatt_parallel_2019}
\BIBentryALTinterwordspacing
C.~Figgatt, A.~Ostrander, N.~M. Linke, K.~A. Landsman, D.~Zhu, D.~Maslov, and
  C.~Monroe, ``\BIBforeignlanguage{en}{Parallel entangling operations on a
  universal ion-trap quantum computer},''
  \emph{\BIBforeignlanguage{en}{Nature}}, vol. 572, no. 7769, pp. 368--372,
  Aug. 2019. [Online]. Available:
  \url{https://www.nature.com/articles/s41586-019-1427-5}
\BIBentrySTDinterwordspacing

\bibitem{de_clercq_parallel_2016}
\BIBentryALTinterwordspacing
L.~E. de~Clercq, H.-Y. Lo, M.~Marinelli, D.~Nadlinger, R.~Oswald,
  V.~Negnevitsky, D.~Kienzler, B.~Keitch, and J.~P. Home, ``Parallel
  {Transport} {Quantum} {Logic} {Gates} with {Trapped} {Ions},'' \emph{Physical
  Review Letters}, vol. 116, no.~8, p. 080502, Feb. 2016, publisher: American
  Physical Society. [Online]. Available:
  \url{https://link.aps.org/doi/10.1103/PhysRevLett.116.080502}
\BIBentrySTDinterwordspacing

\bibitem{pino_demonstration_2021}
\BIBentryALTinterwordspacing
J.~M. Pino, J.~M. Dreiling, C.~Figgatt, J.~P. Gaebler, S.~A. Moses, M.~S.
  Allman, C.~H. Baldwin, M.~Foss-Feig, D.~Hayes, K.~Mayer, C.~Ryan-Anderson,
  and B.~Neyenhuis, ``\BIBforeignlanguage{en}{Demonstration of the trapped-ion
  quantum {CCD} computer architecture},''
  \emph{\BIBforeignlanguage{en}{Nature}}, vol. 592, no. 7853, pp. 209--213,
  Apr. 2021. [Online]. Available:
  \url{https://www.nature.com/articles/s41586-021-03318-4}
\BIBentrySTDinterwordspacing

\bibitem{fallek_transport_2016}
\BIBentryALTinterwordspacing
S.~D. Fallek, C.~D. Herold, B.~J. McMahon, K.~M. Maller, K.~R. Brown, and J.~M.
  Amini, ``\BIBforeignlanguage{en}{Transport implementation of the
  {Bernstein}–{Vazirani} algorithm with ion qubits},''
  \emph{\BIBforeignlanguage{en}{New Journal of Physics}}, vol.~18, no.~8, p.
  083030, Aug. 2016, publisher: IOP Publishing. [Online]. Available:
  \url{https://doi.org/10.1088/1367-2630/18/8/083030}
\BIBentrySTDinterwordspacing

\bibitem{nguyen_extending_2020}
\BIBentryALTinterwordspacing
T.~Nguyen, A.~Santana, T.~Kharazi, D.~Claudino, H.~Finkel, and A.~McCaskey,
  ``Extending {C}++ for {Heterogeneous} {Quantum}-{Classical} {Computing},''
  \emph{arXiv:2010.03935 [quant-ph]}, Oct. 2020, arXiv: 2010.03935. [Online].
  Available: \url{http://arxiv.org/abs/2010.03935}
\BIBentrySTDinterwordspacing

\bibitem{qcor_website}
\BIBentryALTinterwordspacing
{Oak Ridge National Laboratory}, ``{QCOR Website}.'' [Online]. Available:
  \url{https://qcor.ornl.gov/}
\BIBentrySTDinterwordspacing

\bibitem{lao_designing_2021}
L.~Lao, P.~Murali, M.~Martonosi, and D.~Browne, ``Designing {Calibration} and
  {Expressivity}-{Efficient} {Instruction} {Sets} for {Quantum} {Computing},''
  in \emph{2021 {ACM}/{IEEE} 48th {Annual} {International} {Symposium} on
  {Computer} {Architecture} ({ISCA})}, Jun. 2021, pp. 846--859.

\bibitem{rajakumar_generating_2020}
\BIBentryALTinterwordspacing
J.~Rajakumar, J.~Moondra, S.~Gupta, and C.~D. Herold, ``Generating {Target}
  {Graph} {Couplings} for {QAOA} from {Native} {Quantum} {Hardware}
  {Couplings},'' \emph{arXiv:2011.08165 [physics, physics:quant-ph]}, Nov.
  2020, arXiv: 2011.08165. [Online]. Available:
  \url{http://arxiv.org/abs/2011.08165}
\BIBentrySTDinterwordspacing

\bibitem{nam_automated_2018}
\BIBentryALTinterwordspacing
Y.~Nam, N.~J. Ross, Y.~Su, A.~M. Childs, and D.~Maslov,
  ``\BIBforeignlanguage{en}{Automated optimization of large quantum circuits
  with continuous parameters},'' \emph{\BIBforeignlanguage{en}{npj Quantum
  Information}}, vol.~4, no.~1, pp. 1--12, May 2018, number: 1 Publisher:
  Nature Publishing Group. [Online]. Available:
  \url{https://www.nature.com/articles/s41534-018-0072-4}
\BIBentrySTDinterwordspacing

\bibitem{amy_staqfull-stack_2020}
\BIBentryALTinterwordspacing
M.~Amy and V.~Gheorghiu, ``\BIBforeignlanguage{en}{staq—{A} full-stack
  quantum processing toolkit},'' \emph{\BIBforeignlanguage{en}{Quantum Science
  and Technology}}, vol.~5, no.~3, p. 034016, Jun. 2020, publisher: IOP
  Publishing. [Online]. Available:
  \url{https://doi.org/10.1088/2058-9565/ab9359}
\BIBentrySTDinterwordspacing

\bibitem{figgatt_complete_2017}
\BIBentryALTinterwordspacing
C.~Figgatt, D.~Maslov, K.~A. Landsman, N.~M. Linke, S.~Debnath, and C.~Monroe,
  ``\BIBforeignlanguage{en}{Complete 3-{Qubit} {Grover} search on a
  programmable quantum computer},'' \emph{\BIBforeignlanguage{en}{Nature
  Communications}}, vol.~8, no.~1, p. 1918, Dec. 2017. [Online]. Available:
  \url{https://www.nature.com/articles/s41467-017-01904-7}
\BIBentrySTDinterwordspacing

\bibitem{dumitrescu_cloud_2018}
\BIBentryALTinterwordspacing
E.~Dumitrescu, A.~McCaskey, G.~Hagen, G.~Jansen, T.~Morris, T.~Papenbrock,
  R.~Pooser, D.~Dean, and P.~Lougovski, ``Cloud {Quantum} {Computing} of an
  {Atomic} {Nucleus},'' \emph{Physical Review Letters}, vol. 120, no.~21, p.
  210501, May 2018, publisher: American Physical Society. [Online]. Available:
  \url{https://link.aps.org/doi/10.1103/PhysRevLett.120.210501}
\BIBentrySTDinterwordspacing

\bibitem{gheorghiu_quantum_2018}
\BIBentryALTinterwordspacing
V.~Gheorghiu, ``\BIBforeignlanguage{en}{Quantum++: {A} modern {C}++ quantum
  computing library},'' \emph{\BIBforeignlanguage{en}{PLOS ONE}}, vol.~13,
  no.~12, p. e0208073, Dec. 2018, publisher: Public Library of Science.
  [Online]. Available:
  \url{https://journals.plos.org/plosone/article?id=10.1371/journal.pone.0208073}
\BIBentrySTDinterwordspacing

\bibitem{morrison_just_2020}
B.~C.~A. Morrison, A.~J. Landahl, D.~S. Lobser, K.~M. Rudinger, A.~E. Russo,
  J.~W. Van Der~Wall, and P.~Maunz, ``Just {Another} {Quantum} {Assembly}
  {Language} ({Jaqal}),'' in \emph{2020 {IEEE} {International} {Conference} on
  {Quantum} {Computing} and {Engineering} ({QCE})}, Oct. 2020, pp. 402--408.

\bibitem{martinez_compiling_2016}
\BIBentryALTinterwordspacing
E.~A. Martinez, T.~Monz, D.~Nigg, P.~Schindler, and R.~Blatt,
  ``\BIBforeignlanguage{en}{Compiling quantum algorithms for architectures with
  multi-qubit gates},'' \emph{\BIBforeignlanguage{en}{New Journal of Physics}},
  vol.~18, no.~6, p. 063029, Jun. 2016. [Online]. Available:
  \url{https://iopscience.iop.org/article/10.1088/1367-2630/18/6/063029}
\BIBentrySTDinterwordspacing

\bibitem{younis_qfast_2021}
\BIBentryALTinterwordspacing
E.~Younis, K.~Sen, K.~Yelick, and C.~Iancu, ``{QFAST}: {Conflating} {Search}
  and {Numerical} {Optimization} for {Scalable} {Quantum} {Circuit}
  {Synthesis},'' \emph{arXiv:2103.07093 [quant-ph]}, Mar. 2021, arXiv:
  2103.07093. [Online]. Available: \url{http://arxiv.org/abs/2103.07093}
\BIBentrySTDinterwordspacing

\bibitem{wu_qgo_2021}
\BIBentryALTinterwordspacing
X.-C. Wu, M.~G. Davis, F.~T. Chong, and C.~Iancu, ``{QGo}: {Scalable} {Quantum}
  {Circuit} {Optimization} {Using} {Automated} {Synthesis},''
  \emph{arXiv:2012.09835 [quant-ph]}, Jul. 2021, arXiv: 2012.09835. [Online].
  Available: \url{http://arxiv.org/abs/2012.09835}
\BIBentrySTDinterwordspacing

\bibitem{lu_global_2019}
\BIBentryALTinterwordspacing
Y.~Lu, S.~Zhang, K.~Zhang, W.~Chen, Y.~Shen, J.~Zhang, J.-N. Zhang, and K.~Kim,
  ``\BIBforeignlanguage{en}{Global entangling gates on arbitrary ion qubits},''
  \emph{\BIBforeignlanguage{en}{Nature}}, vol. 572, no. 7769, pp. 363--367,
  Aug. 2019. [Online]. Available:
  \url{http://www.nature.com/articles/s41586-019-1428-4}
\BIBentrySTDinterwordspacing

\bibitem{wu_tilt_2021}
X.-C. Wu, D.~M. Debroy, Y.~Ding, J.~M. Baker, Y.~Alexeev, K.~R. Brown, and
  F.~T. Chong, ``{TILT}: {Achieving} {Higher} {Fidelity} on a {Trapped}-{Ion}
  {Linear}-{Tape} {Quantum} {Computing} {Architecture},'' in \emph{2021 {IEEE}
  {International} {Symposium} on {High}-{Performance} {Computer} {Architecture}
  ({HPCA})}, Feb. 2021, pp. 153--166, iSSN: 2378-203X.

\bibitem{kielpinski_architecture_2002}
\BIBentryALTinterwordspacing
D.~Kielpinski, C.~Monroe, and D.~J. Wineland,
  ``\BIBforeignlanguage{en}{Architecture for a large-scale ion-trap quantum
  computer},'' \emph{\BIBforeignlanguage{en}{Nature}}, vol. 417, no. 6890, pp.
  709--711, Jun. 2002, bandiera\_abtest: a Cg\_type: Nature Research Journals
  Number: 6890 Primary\_atype: Reviews Publisher: Nature Publishing Group.
  [Online]. Available: \url{https://www.nature.com/articles/nature00784}
\BIBentrySTDinterwordspacing

\bibitem{young:2019:rg-exp-insights}
J.~S. Young, J.~Riedy, T.~M. Conte, V.~Sarkar, P.~Chatarasi, and S.~Srikanth,
  ``Experimental insights from the rogues gallery,'' in \emph{2019 IEEE
  International Conference on Rebooting Computing (ICRC)}, Nov 2019, pp. 1--8.

\end{thebibliography}
